\documentclass[aps,prb,twocolumn,showpacs,preprintnumbers,amsmath,amssymb,superscriptaddress]{revtex4}

\usepackage{graphicx}
\usepackage{dcolumn}
\usepackage{bm}
\usepackage{color}
\usepackage{amsmath}
\usepackage{amssymb}

\begin{document}

\title{The role of the indirect tunneling processes and asymmetry in couplings in orbital Kondo
transport through double quantum dots}

\author{Piotr Trocha}
\email{ptrocha@amu.edu.pl}\affiliation{Department of Physics,
Adam Mickiewicz University, 61-614 Pozna\'n, Poland}

\date{\today}

\begin{abstract}
System of two quantum dots attached to external electrodes is
considered theoretically in orbital Kondo regime. In general, the
double dot system is coupled via both Coulomb interaction and
direct hoping. Moreover, the indirect hopping processes between
the dots (through the leads) are also taken into account. To
investigate system's electronic properties we apply slave-boson
mean field (SBMF) technique. With help of the SBMF approach the
local density of states for both dots and the transmission
(as well as linear and differencial conductance) is calculated. We show that
Dicke- and Fano-like line shape may emerge in transport
characteristics of the double dot system. Moreover, we observed
that these modified Kondo resonances are very susceptible to the
change of the indirect coupling's strength. We have also shown
that the Kondo temperature become suppressed with increasing
asymmetry in the dot-lead couplings when there is no indirect
coupling. Moreover, when the indirect coupling is turned on the
Kondo temperature becomes suppressed. 
By allowing a relative sign of the nondiagonal elements of the 
coupling matrix with left and right electrode, we extend our 
investigations become more generic.
Finally, we have also included the level renormalization 
effects due to indirect tunneling, which in most papers is not taken into account.

\end{abstract}




\maketitle
\section{Introduction}\label{Sec:1}


Originally, Kondo effect was discovered in non-magnetic
metal containing magnetic impurities at low temperature. The
effect comes from strong electron correlations and can be regarded
as interactions of the impurity spin with cloud of the conduction
electrons in metal. Scattering of the conduction electrons from
the localized magnetic impurities leads to increase of the
resistivity at low temperature.~\cite{Kondo} In recent two decades
one could notice revival of this effect as it was predicted and
observed in transport through quantum
dots.~\cite{cronewett,gores,glazman} Although, in this case Kondo
resonance leads to increasing conductivity (not resistivity as in
original Kondo effect) with decreasing temperature below so-called
Kondo temperature $T_K$, the physical mechanism of the phenomena
is common. Here, the role of the magnetic impurity plays a spin on
the dot.

\begin{figure}[t]
\begin{center}
  \includegraphics[width=0.6\columnwidth]{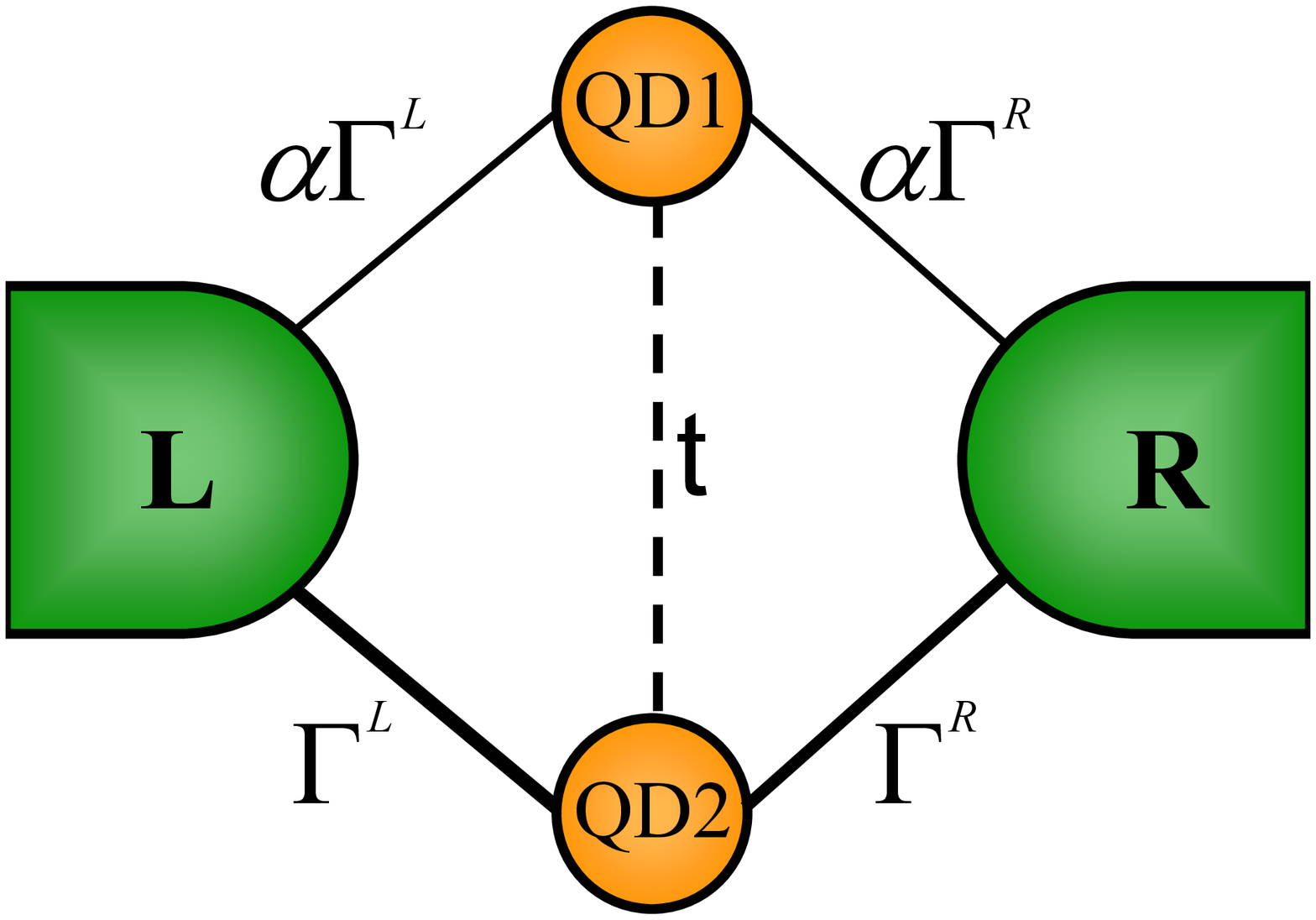}
  \caption{\label{Fig:1}
  Schematic picture of the double dot system. The parameter $\alpha$ takes into account difference in the coupling
of the two dots to external leads ($\alpha\in\langle 0,1\rangle$).
Tuning parameter $\alpha$ one can change geometry of the system
from the parallel one for $\alpha=1$ to the T-shaped geometry for
$\alpha=0$.}
\end{center}
\end{figure}

However, despite of regarding spin degree of freedom one may assume any two-valued quantum numbers, as for instance, an orbital degree of freedom, to realize Kondo effect. In the case of (at least) two discrete orbital levels coupled to the external leads one may deal with orbital Kondo effect.~\cite{zawadowski,boese,imry}
To explain mechanism of creation of the Kondo state let us
introduce spinless electrons in a system of two single-level quantum
dots coupled to external leads as shown in Fig.~\ref{Fig:1}.
The dots' energy levels are well below the
Fermi level of the leads. Due to large interdot Coulomb repulsion only one of them is occupied by an electron. On the other hand, removing an existing electron on the double dot requires adding energy to the system. Thus, the system is in the deep Coulomb blockade and the sequential tunneling events are prohibited. However, according to
the Heisenberg uncertainty principle, the higher-order processes, however,
may appear on a very short time scale. Assuming that initially the upper dot was occupied, then it can tunnel onto the Fermi level of the lead (left or right) and simultaneously another electron from the
Fermi level of the left or right electrode may tunnel to the lower dot. As a result, the charge exchange occured between the dots.
A coherent superpostion of
such coherent events give rise to the sharp resonance in the density of
states at the Fermi level.

Altough orbital Kondo
effect has been investigated in different geometries of the two orbital level system both experimentally~\cite{wilhelm,hubel} and theoretically~\cite{sunG,sztenkiel,sztenkiel2,holle,krychowski,schon} there is a little comprehensive studies on the influence of the indirect coupling and/or asymmetry couplings on this phenomenon. In the case of spin Kondo effect most of researchers assume the maximal value of this coupling.~\cite{vernekPE,delftPRB05,lopezPRB05,konikPRL07,zhangPRB05,ding,tanakaPRB} However, the maximal value of the indirect coupling in the case of the orbital Kondo effect may leads to suppression of Kondo resonance. Specifically, this effect is destroyed when the magnetic flux achieves $2n\pi$ ($n\in\mathbb{Z}$)~\cite{wen,kubo08} due to the formation of the bound state in the continuum (BIC)~\cite{bic}. Recently, Kubo {\it et al.} have investigated both spin and orbital Kondo effect in double quantum dot regarding non-maximal values of the indirect coupling~\cite{kubo08,kubo11}. They have found that in the condition of intermediate indirect coupling strength the differential conductance reveals two
kinds of peaks.
Altough, we considered similar system, the role of this paper is to give more insight into the connection between the role of the indirect coupling and various interference effects. Moreover, we assume different asymmetry coupling of the dots to the external leads. More specifically, in the system considered here one dot is coupled to the leads with constant strength, whereas the coupling of the second dot to the leads can be continuously tuned. Time reversal symmetry implies that the amplitude of the indirect coupling strength may change sign~\cite{gurvitz}. Thus, we also include this case in our consideration.

In very recent experiment Tarucha {\it et al.}~\cite{taruchaPRL11} observed that the period of the Aharonov-Bohm oscillations are halved and the phase changes by half a period for the antibonding state from those of the bonding. They conclude that these features can be related to the indirect interdot coupling via the two electrodes.

We also point out that similar double dot system has been investigated in Refs.~[\onlinecite{medenPRL06,medenPRL09}] where the authors have found a novel pair of
correlation-induced resonances as well as they studied the charging of a narrow QD level capacitively coupled
to a broad one. However, the correlation-induced effects are not related to the Kondo physics.


As we mentioned before various quantum interference effects, which
were previously reported in  atomic physics or quantum optics~\cite{dicke1,dicke2,Fano}, were
also discovered in electronic transmission through QDs systems
attached to the leads\cite{guevaraD,shahbazyan,brandes,trocha,trocha1}. Here, we consider
Dicke and Fano resonances. In original Dicke effect one observes
in spontaneous emission spectra a strong and very narrow resonance
which coexist with much broader line. This occurs when the
distance between the atoms is much smaller than the wavelength of
the emitted light (by individual atom). The former resonance,
associated with a state which is weakly coupled to the
electromagnetic field, is called subradiant mode, and the latter,
strongly coupled to the electromagnetic field refers to
superradiant mode. In the case of electronic transport in
mesoscopic systems (for instance, quantum dots) this effect is due
to indirect coupling of QDs through the leads. Generally, indirect
coupling can lead to the formation of the bonding and antibonding
states. As a result, a broad peak corresponding to the bonding
state and a narrow one referring to the antibonding state emerge
in the density of states~\cite{wunsch,trochaJNN}.

Let us now tell something about Fano effect. In experiment, Fano
effect manifests itself as asymmetric line shape in emission
spectra. It comes from quantum interference of waves resonantly
transmitted through a discrete level and those transmitted
nonresonantly through a continuum of states. The effect was
observed in optics and also in electronic transport through QDs
systems \cite{Gores2,clerk,kobayashi,johnson,sasaki}. However, in this
case the Fano phenomenon is  due to the quantum interference of
electron waves transmitted coherently through the dot and those
transmitted directly between the leads~\cite{Bulka}.

It is convenient to associate resonant channel with discrete level
and nonresonant channel with continuum of states. When the
electron wave passes through resonant channel its phase changes by
$\pi$ (within $\Gamma$), whereas the phase of electron waves in
nonresonant channel changes very slowly around the resonant level
($\Gamma$ is the width of the discrete level). (Of course, Fano
effect occurs only when discrete level is embedded into a
continuum.) Consequently, on the one side of the discrete level
electron waves through two channels interfere constructively,
whereas on the other side they interfere destructively. As a
result, one observes asymmetric line in conductance around the
discrete level position.

This effect can be also observed in system of two quantum dots
embedded in two arms of AB
ring\cite{guevara,lu05,ding,chi,trocha2} or in the so-called $T$
geometry~\cite{sasaki,zitkoPRB10}. Here, very narrow (broad)
level, which is weakly (strongly) coupled to the leads,
corresponds to the resonant (nonresonant) channel. The narrow
level must appear within the broad one. As mentioned before the
difference in the coupling strengths of the bonding and
antibonding states is due to indirect coupling. The phase shift of
wave function in the broad level is negligible when the energy
changes within the narrow level and Fano resonance may appear.

In this paper the orbital Kondo effect in electronic transport
through two coupled quantum dots is considered theoretically.
Generally, the quantum dots may interact via both Coulomb
repulsion and hopping term. To calculate local density of states
(LDOS) for both dots, transmission, and differential conductance  we employ
slave-boson mean field approach. To show  the formation of the bound state in the continuum as the indirect coupling strength approaches its maximal value we calculate the Friedel phase. Due to emergence of the BIC the Fiedel phase, usually continuous, changes abruptly at the energy corresponding to the BIC. Finally, to be more familiar with experiment we show differential conductance.

The paper is organized as follows. In Section 2 we describe the
model of a double-dot system which is taken under considerations.
We also present there the slave-boson mean-field technique used to
calculate the basic transport characteristics. Numerical results
on the orbital Kondo problem are shown and discussed in Section 3.
Final conclusions are presented in Section 4.



\section{Theoretical description}\label{Sec:2}


\subsection{Model}\label{sec:2A}
%
To investigate various interference effects in Kondo regime we
consider (spinless) Anderson Hamiltonian for double quantum dots
coupled to external leads. Experimentally, such system may be
realized by applying a magnetic field which lifts spin degeneracy
of each dot. Generally, this Hamiltonian consists of three parts,
\begin{equation}\label{Eq:Hamiltonian}
\hat{H}=\hat{H}_{\rm c} +\hat{H}_{\rm DQD}+\hat{H}_{\rm T},
\end{equation}
where the first term, $\hat{H}_{\rm c}$, describes nonmagnetic
electrodes in the non-interacting quasi-particle approximation,
$\hat{H}_{\rm c}=\hat{H}_{L}+\hat{H}_{R}$, with $\hat{H}_{\beta
}=\sum_{\mathbf{k}}
\varepsilon_{\mathbf{k}\beta}c^{\dagger}_{\mathbf{k}\beta}
c_{\mathbf{k}\beta}$ (for electrodes $\beta ={\rm L,R}$). Here,
$c^{\dagger}_{\mathbf{k}\beta}$ ($c_{\mathbf{k}\beta}$) creates
(annihilates) an electron with the wave vector $\mathbf{k}$ in the
lead $\beta$, whereas $\varepsilon_{\mathbf{k}\beta}$ denotes the
corresponding single-particle energy.

The next term of Hamiltonian (\ref{Eq:Hamiltonian}) describes two
coupled quantum dots,
\begin{equation}
   \hat{H}_{DQD}=\sum_{i}\limits\varepsilon_{i}d^\dag_{i}d_{i}+
   t(d^\dag_{1}d_{2}+h.c.)+
   Un_{1}n_{2},
\end{equation}
where $n_{i}=d^\dag_{i}d_{i}$ is the particle number operator,
$\varepsilon_{i}$ is the discrete energy level of the $i$-th dot
($i=1,2$), $t$ denotes the inter-dot hopping parameter (assumed
real), whereas $U$ is the inter-dot Coulomb integral.

The last term, $H_{\rm T}$, of Hamiltonian (\ref{Eq:Hamiltonian})
describes electron tunneling between the leads and dots, and takes
the form
\begin{equation}
   \hat{H}_{\rm T}=\sum_{\mathbf{k}\beta}\limits\sum_{i=1,2}
   \limits (V_{i\mathbf{k}}^\beta c^\dag_{\mathbf{k}\beta}d_{i}+\rm
   h.c.),
   \end{equation}
where $V_{i\mathbf{k}}^\beta$ are the relevant matrix elements.

Finite widths of the discrete dots' energy levels come from
coupling to the external leads and may be expressed in the form
$\Gamma^\beta_{ii}(\varepsilon)=2\pi |V_{i\mathbf{k}}^\beta|^2 \rho$, where
$\rho$ denotes density of states in the left and right lead
($\rho_L=\rho_R\equiv\rho$). Furthermore, we assume that
$\Gamma^\beta_{ii}$ is constant within the electron band,
$\Gamma^\beta_{ii}(\varepsilon)=\Gamma^\beta_{ii}={\rm const}$ for
$\varepsilon\in\langle-D,D\rangle$, and
$\Gamma^\beta_{ii}(\varepsilon)=0$ otherwise. Here, $2D$ denotes the
electron band width.

For the system taken under considerations, the dot-lead couplings
can be written in a matrix form,
\begin{equation}\label{Eq:coupling}
  \mathbf{\Gamma}^\beta = \left(
\begin{array}{cc}
  \Gamma^\beta_{11} & \Gamma^\beta_{12} \\
  \Gamma^\beta_{21} & \Gamma^\beta_{22}
\end{array}
\right),
\end{equation}
where the off-diagonal matrix elements are assumed to be
$\Gamma^\beta_{12} = \Gamma^\beta_{21} = q_\beta
\sqrt{\Gamma^\beta_{11} \Gamma^\beta_{22}}$.\cite{shahbazyan,kuboPRB06} The off-diagonal
matrix elements of $\mathbf{\Gamma}^\beta$ take into account
various interference effects resulting from indirect tunneling
processes between two quantum dots via the leads. These
off-diagonal matrix elements may be significantly reduced in
comparison to the diagonal matrix elements $\Gamma^\beta_{ii}$ or
even totally suppressed due to complete destructive interference.
To take all those effects into account, the parameters $q_{\rm L}$
and $q_{\rm R}$ are introduced. Furthermore, we assume that
$q_\beta$ are real numbers and obey the condition
$|q_\beta| \le 1$. When $q_{\beta}$ is nonzero, the processes in
which an electron tunnels from one dot to the $\beta$-lead and
then (coherently) to the another dot are allowed. Introducing
parameter $\alpha$, which takes into account difference in the
coupling of the two dots to external leads, the coupling matrix
Eq.~(\ref{Eq:coupling}) can be rewritten in the form,
\begin{equation}\label{Eq:coupling}
  \mathbf{\Gamma}^\beta = \Gamma^\beta\left(
\begin{array}{cc}
  \alpha & q_\beta\sqrt{\alpha} \\
  q_\beta\sqrt{\alpha} & 1
\end{array}
\right).
\end{equation}
Tuning parameter $\alpha$ one can change geometry of the system
from the parallel one for $\alpha=1$ to the T-shaped geometry for
$\alpha=0$. In the T-shaped geometry upper dot is disconnected
from the leads. All intermediate values of the $\alpha$ refer to
an intermediate geometry where each of the two dots is coupled to
leads with different strength. We further assume symmetric
coupling $\Gamma^{\rm L} = \Gamma^{\rm R}\equiv \Gamma/2$, where
$\Gamma$ is energy unit.

\subsection{Method}\label{sec:2B}
We perform our calculations in large interdot charging energy
limit, more specifically, when $U\rightarrow\infty$. Slave boson
approach is one of the techniques which allows investigate
strongly correlated fermions in low temperatures\cite{coleman}.
This method relies on introducing auxiliary operators for the dots
and replacing of the dots' creation and annihilation operators by
$f^{\dag}_{i}b$, $b^{\dag}f_{i}$ respectively. Here, slave-boson operator $b^{\dag}$
creates an empty state, whereas pseudo-fermion operator $f^{\dag}_{i}$ creates singly
occupied state with an electron in the i-th dot. To eliminate
non-physical states, the following constraint has to be imposed on
the new quasi-particles,
\begin{equation}\label{Eq:constraint}
    Q=\sum_{i}\limits
    f^{\dag}_{i}f_{i}+b^{\dag}b=1.
\end{equation}
Above constraint prevents double occupancy of the dots; dots are
empty or singly occupied.

In the next step the Hamiltonian (\ref{Eq:Hamiltonian}) of the
system is replaced by an effective Hamiltonian, expressed in terms
of the auxiliary boson $b$ and pseudo-fermion $f_{i}$ operators as,
\begin{eqnarray}\label{Eq:NewHamiltonian}
\tilde{H}=\sum_{\mathbf k\beta} \varepsilon_{{\mathbf
k}\beta}c^{\dagger}_{{\mathbf k}\beta} c_{{\mathbf k}\beta}+
\sum_{i}\limits\varepsilon_{i}f^\dag_{i}f_{i} + (t
f^\dag_{1}bb^{\dag}f_{2} +\rm {\rm H.c.})
  \nonumber \\
   + \sum_{{\mathbf k}\beta}\sum_{i}\limits
   (V_{i{\mathbf k}}^\beta
   c^\dag_{{\mathbf k}\beta}b^{\dag}f_{i}+{\rm H.c.})
   +\lambda\left(\sum_{i}\limits
    f^{\dag}_{i}f_{i}+b^{\dag}b-1\right).
   \end{eqnarray}
To avoid double occupancy of the DQD system the constriction
condition (Eq.~(\ref{Eq:constraint})) has been incorporated in
Hamiltonian (\ref{Eq:NewHamiltonian}) by introducing the term with
the Lagrange multiplier $\lambda$.

However, after such transformation our Hamiltonian is still rather
complex and hard to solve. To get rid of this problem we apply
mean field (MF) approximation in which the boson field $b$ is replaced
by a real and independent of time $c$ number,
$b(t)\rightarrow\langle b(t)\rangle\equiv{\tilde b}$. This
approximation neglects fluctuations around the average value
$\langle b(t)\rangle$ of the slave boson operator, but is
sufficient to describe correctly those leading to the Kondo
effect. It also restricts our considerations to the low bias
regime ($eV\ll |\varepsilon_i|$).


With the following definitions of the renormalized parameters:
$\tilde{t}=t\tilde{b}^2$, $\tilde{V}_{i\mathbf {k}}^\beta
=V_{i\mathbf {k}}^\beta\tilde{b}$ and
$\tilde{\varepsilon}_{i}=\varepsilon_{i}+\lambda$, one can rewrite the
effective MF Hamiltonian in the form,
\begin{eqnarray}
\tilde{H}^{MF}=\sum_{\mathbf k\beta} \varepsilon_{{\mathbf
k}\beta}c^{\dagger}_{{\mathbf k}\beta} c_{{\mathbf k}\beta }+
\sum_{i}\limits\tilde{\varepsilon}_{i}f^\dag_{i}f_{i}
   +(\tilde{t}f^\dag_{1}f_{2}+{\rm h.c.})
  \nonumber \\
   +   \sum_{{\mathbf k}\beta}\sum_{i}\limits
   (\tilde{V}_{i\mathbf{k}}^\beta
   c^\dag_{{\mathbf k}\beta}f_{i}+{\rm h.c.})
  +  \lambda\left(\tilde{b}^2-1\right).
   \end{eqnarray}
The unknown parameters $\tilde{b}$ and $\lambda$ have to be found
self-consistently with the help of the following equations;
\begin{equation}\label{Eq:SC1}
    \tilde{b}^2-i\sum_{\sigma}\limits
    \int\frac{d\varepsilon}{2\pi}\langle\langle
    f_{i}|f_{i}^{\dag}\rangle\rangle^<_\varepsilon =1,
\end{equation}
\begin{eqnarray}\label{Eq:SC2}
-i\sum_{i}\limits\int\frac{d\varepsilon}{2\pi}(\varepsilon-\tilde{\varepsilon}_{i})
\langle\langle f_{i}|f^{\dag}_{i}\rangle\rangle^<_\varepsilon +
\lambda\tilde{b}^2=0,
\end{eqnarray}
where $\langle\langle
    f_{i}|f^{\dag}_{j}\rangle\rangle^<_\varepsilon$
    is the Fourier transform of the
    lesser Green function defined as $G^<_{ij\sigma}(t,t^\prime)\equiv \langle\langle
    f_{i}(t)|f^{\dag}_{j}(t')\rangle\rangle^<=i
    \langle
    f^{\dag}_{j}(t')f_{i}(t)\rangle$.
The above equations have been obtained from the constraints
imposed on the slave boson, Eq.~(\ref{Eq:constraint}), and from
the equation of motion for the slave boson operator. The lesser
Green functions $\langle\langle
f_{i}|f^{\dag}_{i}\rangle\rangle^<_\varepsilon$ as well as
retarded $\langle\langle f_{i}|f^{\dag}_{i}\rangle\rangle^r$
(which also is needed in further calculations) have been
determined from the corresponding equation of motion.

To get insight into system's electronic properties we calculate
local density of states (LDOS) and transmission function. The
local density of states for the i-th dot is defined as;
\begin{equation}\label{Eq:DOS}
   D_i=-\frac{\tilde{b}^2}{\pi}\sum_{\sigma}{\rm Im}\left[G_{ii}^r(\varepsilon)\right],
\end{equation}
whereas transmission through the system is expressed
in the form;
\begin{equation}\label{Eq:DOS}
   T(\varepsilon)={\rm Tr}[\mathbf{G}^a\mathbf{\tilde{\Gamma}}^R\mathbf{G}^r\mathbf{\tilde{\Gamma}}^L],
\end{equation}
In above formula $\mathbf{\tilde{\Gamma}}^{\beta}$ stands for
coupling matrix to the $\beta$-th lead with renormalized
parameters
$\tilde{\Gamma}_{ij}^{\beta}=\tilde{b}^2\Gamma_{ij}^{\beta}$, and
$\mathbf{G}^r$ ($\mathbf{G}^a$) denotes Fourier transforms of the
retarded (advanced) Green functions of the dots. It is worth
noting that transmission probability $T(\varepsilon)$ is directly
related with current and linear conductance. In the zero
temperature limit $T\rightarrow 0$ these quantities are given by
formulas
\begin{equation}\label{Eq:current}
    J=\frac{e}{h}\int_{-eV/2}^{eV/2}d\varepsilon\;
    T(\varepsilon),
\end{equation}
and
\begin{equation}\label{Eq:conductance}
    G_{V\to 0}=\lim_{V\rightarrow 0}\frac{dJ}{dV}=\frac{e^2}{h}
    T(\varepsilon=0).
\end{equation}



\section{Numerical results}\label{Sec:3}

\subsection{Dicke effect}\label{Sec:3A}

In the following numerical calculations we assume equal dot energy
levels, $\varepsilon_{i}=\varepsilon_0$ (for $i =1,2$) ($\varepsilon_0$ is
measured from the Fermi level of the leads in equilibrium,
$\mu_L=\mu_R=0$). Moreover, we set the bare level of the dots at
$\varepsilon_{0}=-3\Gamma$, and the bandwidth is assumed to be
$2D=120\Gamma$. All the energy quantities are expressed in the
units of $\Gamma$. The parameters $q_L$ and $q_R$ are assumed to
be equal $q_L=q_R=q$ if not stated otherwise. Taking into account the above parameters,
the Kondo temperature $T_K$ for the symmetric couplings ($\alpha=1$) and disregarding both direct and indirect couplings, ($t=0$, $q=0$) is estimated to be $T_K\approx 10^{-2}\Gamma$.
%
\begin{figure}[t]
\begin{center}
 \includegraphics[width=0.58\columnwidth]{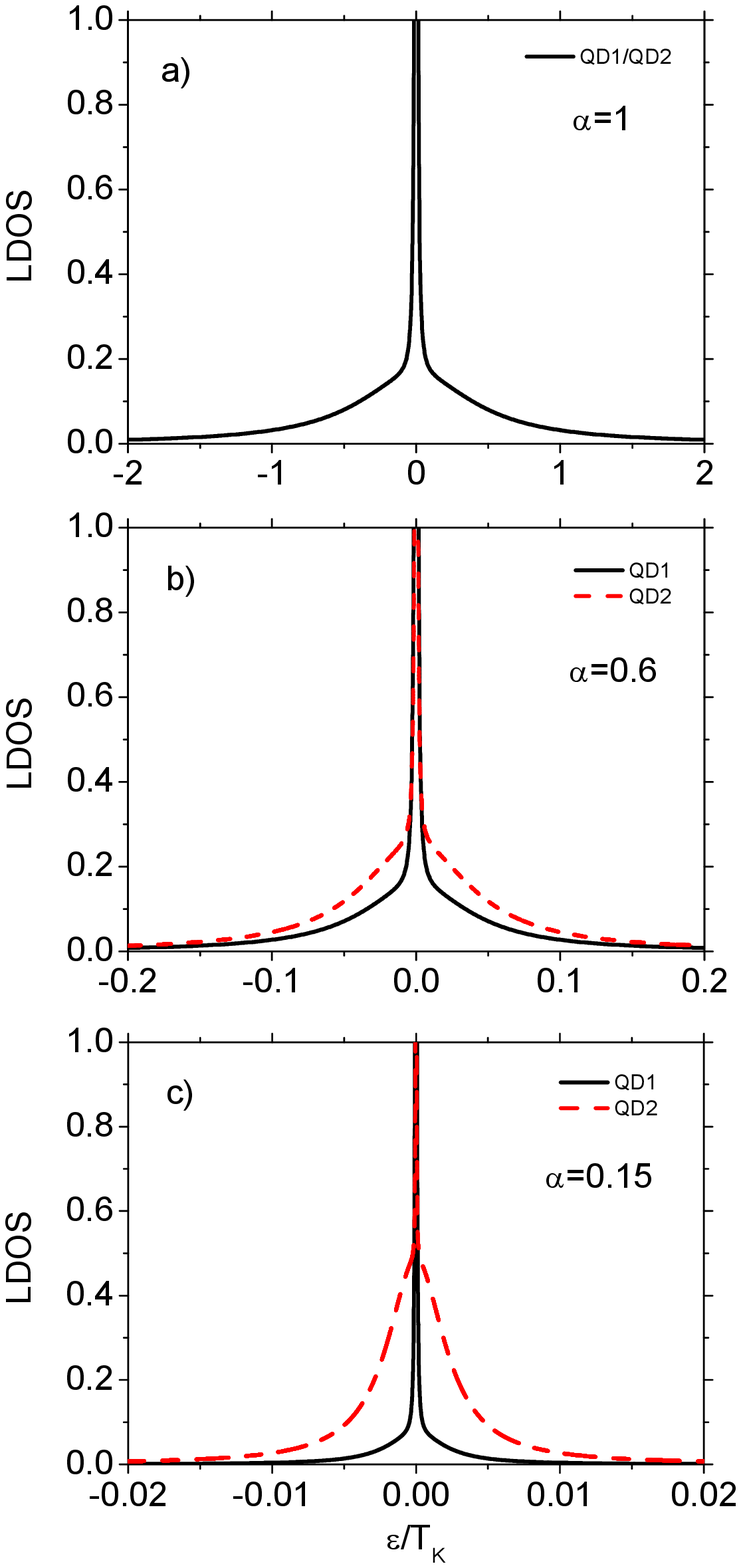}
  \caption{\label{Fig:2}
  Local density of states for the quantum dots QD1 and QD2 obtained for indicated values of $\alpha$ and for
  $q=0.99$. Here, direct hopping between the dots is not allowed, because $t=0$.}
  \end{center}
\end{figure}
In this section we consider the situation when dots are not
directly coupled, which corresponds to the case with vanishing
interdot tunnel coupling parameter $t=0$. On the other side, one
should remember that dots are still coupled (indirectly) through
the leads what is reflected in finite values of the off-diagonal
coupling matrix elements, i.e., $q\neq 0$. In this situation
bonding and antibonding levels, created due to indirect couplings,
coincide. In Fig.~\ref{Fig:2} we show local density of states
for QD1 and QD2 for large off-diagonal matrix couplings ($q=0.99$)
and for different values of the asymmetry parameter $\alpha$.
Firstly, it is clearly presented that the broad and narrow Kondo
peaks in the LDOS are superimposed at energy $\varepsilon=0$ and
the LDOS displays behavior typical for the Dicke effect. By
analogy to the original Dicke phenomenon, one may associate the
narrow (broad) central peak in LDOS with a subradiant
(superradiant) state. The subradiant (superradiant) state
corresponds to longlived (shortlived) state. For symmetric
coupling $\alpha=1$ the LDOS for QD1 and QD2 are the same (see
Fig.~\ref{Fig:2}(a) but when asymmetry appears in couplings
($\alpha\neq 1$) this ceases to be true and the LDOS for both dots
have different line shape. As $\alpha$ drops down the widths of
the Kondo peaks for both dots also diminish. With decreasing
$\alpha$ the broad part of the LDOS for the quantum dot strongly
coupled to the leads (QD2) becomes more pronounced, whereas the
narrow one becomes narrower. The two peaks corresponding to
subradiant and superradiant state are well distinguishable. As a
result, the Dicke effect in LDOS for QD2 is more distinct.
\begin{figure}[t]
\begin{center}
  \includegraphics[width=0.6\columnwidth]{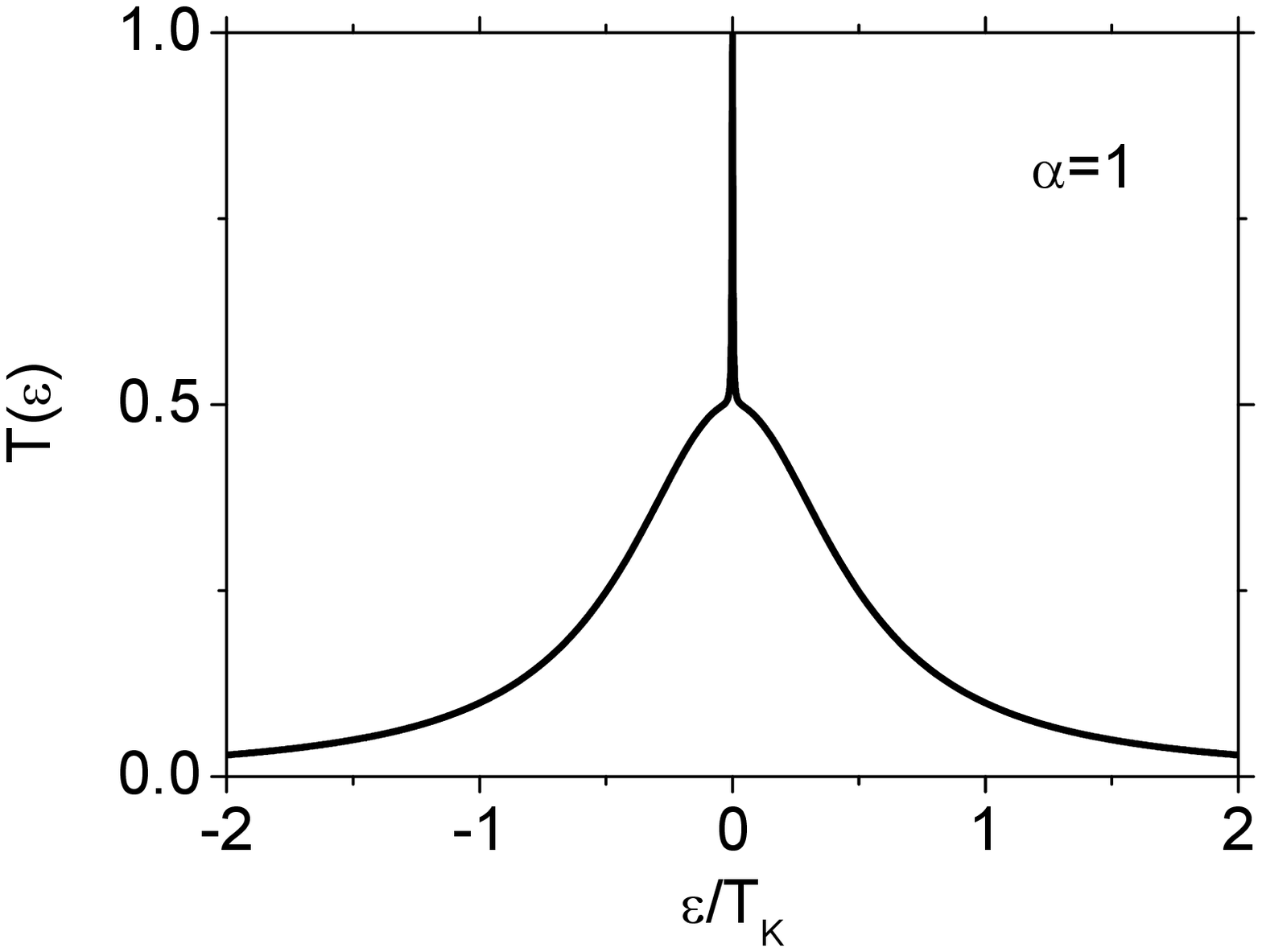}
  \caption{\label{Fig:3}
 The transmission probability calculated for indicated values of $\alpha$ and for
  $q=0.99$ and for $t=0$. The transmission probability has well-defined Dicke line shape.}
  \end{center}
  \end{figure}
  %
The Dicke effect is also noticed in the transmission
shown in Fig.~\ref{Fig:3}. This situation is analogous to that reported in the case of the spin Kondo effect in parallel double dot system.~\cite{trochaJNN} The Dicke
peak appears in the transmission because the phases of the transmission amplitudes for bonding and antibonding channels are equal at zero energy. Thus, the two contributions adds constructively leading to the maximum transmission at $\varepsilon=0$. However, now there is no dip structure at zero energy for $q=1$ (originating from the complete destructive interference), which will be explained further.
Moreover, the effect is
preserved for various values of the asymmetry parameter $\alpha$.
It is worth noting that the linear conductance (see
Eq.~(\ref{Eq:conductance})) reaches unitary limit (two quanta of
$e^2/h$). When $\alpha$ is reduced the effective Kondo temperature
also decreases what can be seen looking at the energy scale in
Figure \ref{Fig:2}.
The origin of this effect comes from the fact
that when $\alpha$ decreases one of the dots becomes detached from
the electrodes. Then, the rate of the higher order tunneling
events (which leads to the Kondo anomaly-see explanation in the
Introduction) also diminishes and finally for $\alpha=0$, when one
of the dot is totally disconnected from the leads, there is no
possibility for such events and no Kondo effect is expected.
\begin{figure}[t]
\begin{center}
  \includegraphics[width=0.82\columnwidth]{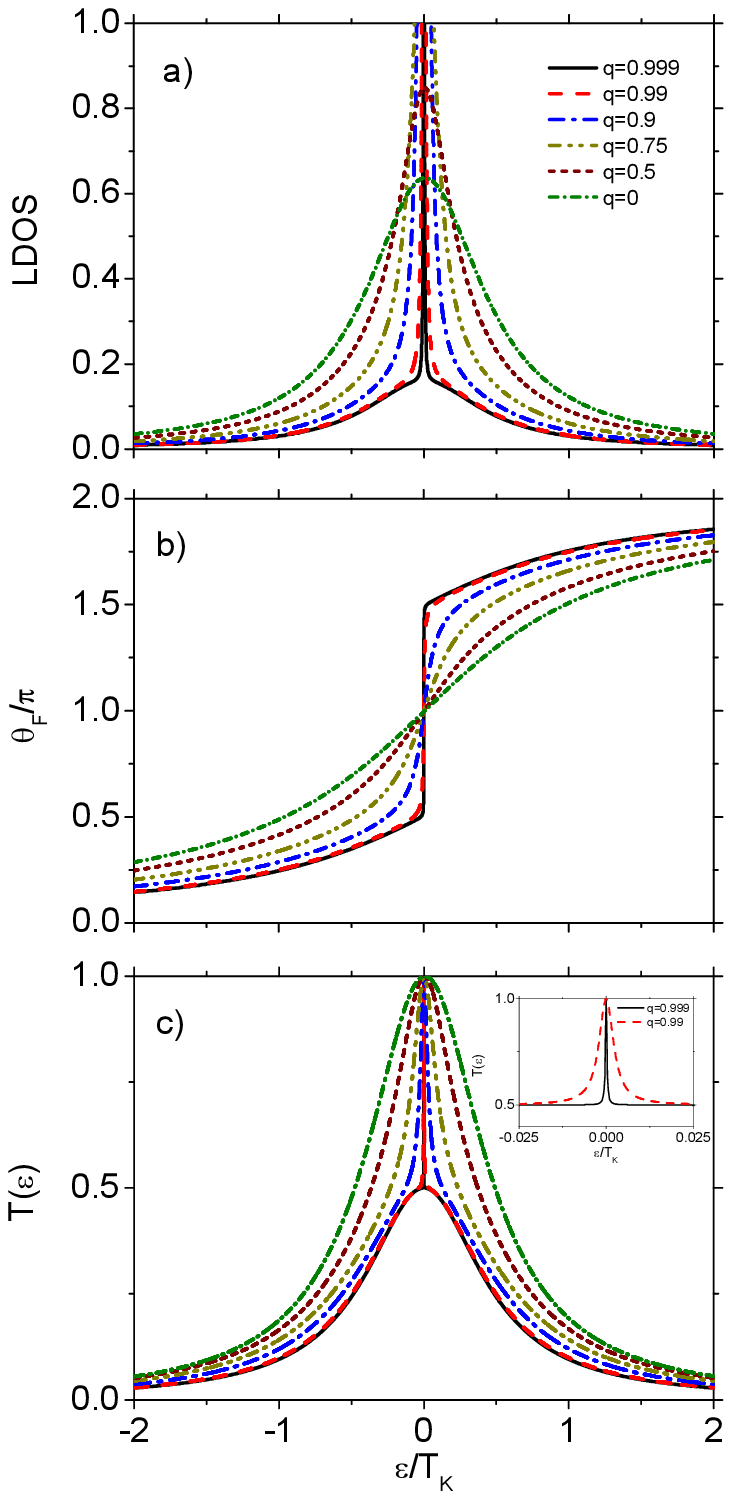}
  \caption{\label{Fig:4}
  Local density of states for the both dots (a) Friedel phase (b) and the transmission (c) calculated for indicated values of the $q$ and for
  $\alpha=1$ and for $t=0$. The well-defined Dicke line shape is only preserved for indirect coupling parameter $q$ close to 1.}
  \end{center}
\end{figure}

As we mentioned in Section \ref{sec:2A} the off-diagonal matrix
elements $\Gamma^\beta_{12}$ may be significantly reduced, so it
is desired to analyze this case. In Fig.~(\ref{Fig:4}) we plotted
the LDOS, Friedel phase and the transmission for various values of
the parameter $q$ which is directly related with amplitude of the
off-diagonal matrix elements. The Friedel phase is related to the LDOS by the following equation:
$d\theta_F/d\varepsilon=\pi D(\varepsilon)$ with $D(\varepsilon)$ being relevant density of states.
One can notice that Dicke effect in
LDOS can be found only when off-diagonal matrix elements are
large, i.e., $q$ close to $1$. With decreasing $q$ the Dicke line
shape is transformed in usual Lorenzian line. Similar behavior is
observed in the transmission (see
Fig.~\ref{Fig:4}(c)). In inset of Fig.~\ref{Fig:4}(c)) we show
that the narrow part of the peak has also well defined shape as
the broad one. Moreover, when $q$ is closer and closer to 1 then
the narrow peak becomes more and more narrower. This is reflected in abrupt (but continuous as long as $q\neq 1$ ) change by $\pi$ of Friedel phase around $\varepsilon=0$ for $q$ close to $1$. It is also found that
the Dicke effect also disappears in transmission when
$q$ is maximal ($q=1$). When $q=1$ the transmission probability
has Lorenzian shape and is described by formula,
\begin{equation}\label{Eq:Tq1}
    T(\varepsilon)=\frac{1}{2}\left(\frac{(1+\alpha)^2\tilde{\Gamma}^2}{(1+\alpha)^2\tilde{\Gamma}^2+
    (\tilde{\varepsilon}_0-\varepsilon)^2}\right),
\end{equation}
where $\tilde{\Gamma}=\tilde{b}^2\Gamma /2$. This equation clearly
shows that no Dicke effect should be expected for $q=1$. This
result resembles that obtained for a noninteracting
system~\cite{shahbazyan}. However, in the Kondo regime the situation is much more complex.
To show this, let us first consider the case with $q=1$ and symmetric couplings, $\alpha=1$. It is well
known that for symmetric ($\alpha=1$) noninteracting ($U=0$)
system as $q$ tends to $1$ one of the peaks becomes progressively
narrowed and finally for $q=1$ a BIC
emerges~\cite{shahbazyan,solisPLA}. As a result the transmission
reveals simple lorenzian lineshape. One may naively believe that
similar situation occurs in the Kondo regime. However, this can
not be true because when the indirect coupling strength is equal
to the dot-lead coupling ($\Gamma_{12}=\Gamma_{11}=\Gamma_{22}$),
the indirect tunneling processes completely destroy coherent
higher-order tunelling events leading to complete suppression of
the Kondo resonance.~\cite{wen} One can also look at this from the
another point of view and explain this as follows: As the
antibonding state becomes a BIC, it is totally decoupled from the
leads and there is no possibility for an electron to exchange
between the two molecular states. Thus, no Kondo effect appears as
the corresponding Kondo temperature is equal to zero.
\begin{figure}[t]
\begin{center}
  \includegraphics[width=0.74\columnwidth]{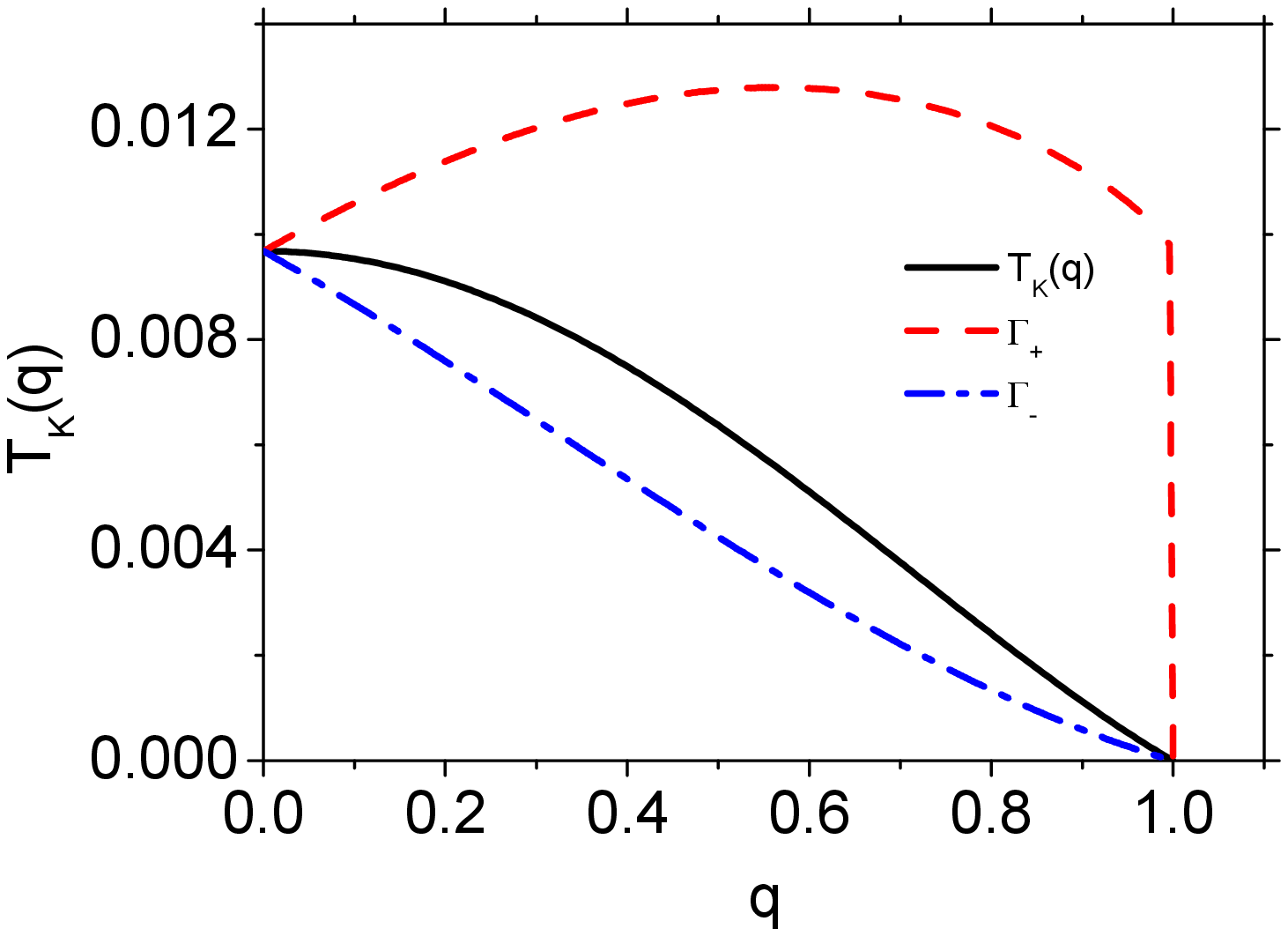}
  \caption{\label{Fig:figTKq}
 The Kondo temperature calculated as a function of parameter $q$ and renormalized width $\tilde{\Gamma}_+$
 ($\tilde{\Gamma}_-$) coresponding to the
 bonding (antybonding) level
 calculated for $\alpha=1$ and for $t=0$.}
  \end{center}
  \end{figure}
%
To support these predictions we calculated the Kondo temperature
for arbitrary value of the parameter $q$. The Kondo temperature
for the symmetric case, $\alpha=1$, acquires the following form
$T_K\equiv\sqrt{\tilde{\varepsilon}_0^2+\tilde{\Gamma}^2(1-q^2)^2}$.
In the deep Kondo regime the renormalized parameter
$\tilde{\varepsilon}_0$ is equal to zero, which is shown in the
Appendix, and thus the above formula clearly show vanishing of the
corresponding Kondo temperature as $q$ tends to $1$. In
Fig.~\ref{Fig:figTKq} the Kondo temperature is displayed as a
function of parameter $q$. For comparison we also plot the $q$
dependence of the renormalized widths of the bonding and
antibonding levels, $\tilde{\Gamma}_b$ and $\tilde{\Gamma}_a^2$,
respectively. In the case $q_L=q_R=q$, the widths acquire the
following form $\tilde{\Gamma}_b=\tilde{b}^2\Gamma_b$,
$\tilde{\Gamma}_a=\tilde{b}^2\Gamma_a$ with
$\Gamma_{b,a}=\frac{1}{2}(\Gamma_{11}+\Gamma_{22})\pm
q\sqrt{\Gamma_{11}\Gamma_{22}}$ and
$\Gamma_{ii}=\Gamma_{ii}^L+\Gamma_{ii}^R$ for $i=1,2$. The
characteristic widths for both distinct channels behave in
different way with varying the strength of the off-diagonal
tunneling processes.
One can notice that the renormalized width for the bonding channel
changes non-monotonically but rather slowly, whereas the one for
the antibonding channel drops monotonically to zero as $q$ reaches maximum
value. This behavior is consistent with the predictions given in
Ref.~\onlinecite{lim} with one exception. The vanishing of
$\tilde{\Gamma}_a^2$ when $q$ is maximal is responsible for the
suppression of the Dicke effect. In contrast to Ref.~\onlinecite{lim} the width $\tilde{\Gamma}_b^2$ drops to zero for $q=1$ due to vanishing of the boson field. We also emphasize that for
$q\in(0,1)$, the $\alpha$ dependances of the renormalized widths
for both channels are different. For large values of the $q$ the
width $\tilde{\Gamma}_a$ changes little with $\alpha$.

\begin{figure}[t]
\begin{center}
  \includegraphics[width=0.74\columnwidth]{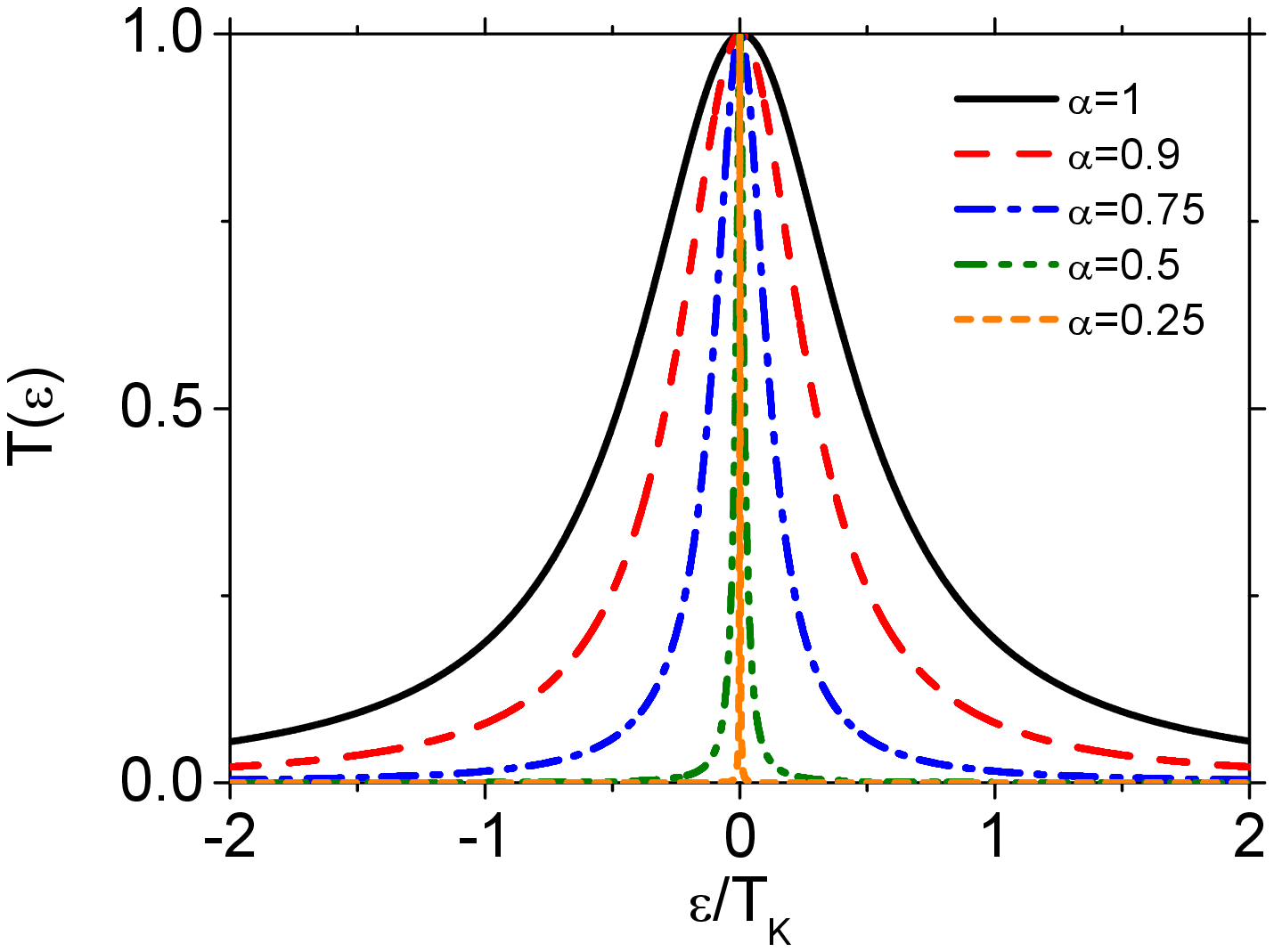}
  \caption{\label{Fig:5}
  Graphical illustration of the Equation \ref{Eq:Tq0}.
  The transmission probability calculated for indicated values of the $\alpha$ and for
  $q=0$.}
  \end{center}
\end{figure}

However, the Kondo effect is also destroyed in more general
case, i.e., for $q=1$ and arbitrary $\alpha$. This can be
understood when one notice that transmission  (\ref{Eq:Tq1})
depends on renormalized parameter $\tilde{b}$ which vanishes in
this case. In the SBMF formalism this is manifested as lack
of solutions of the self-consistent equations~(\ref{Eq:SC1})
and~(\ref{Eq:SC2}) (which then form a contradictory system of
equations).

\begin{figure}[t]
\begin{center}
  \includegraphics[width=0.65\columnwidth]{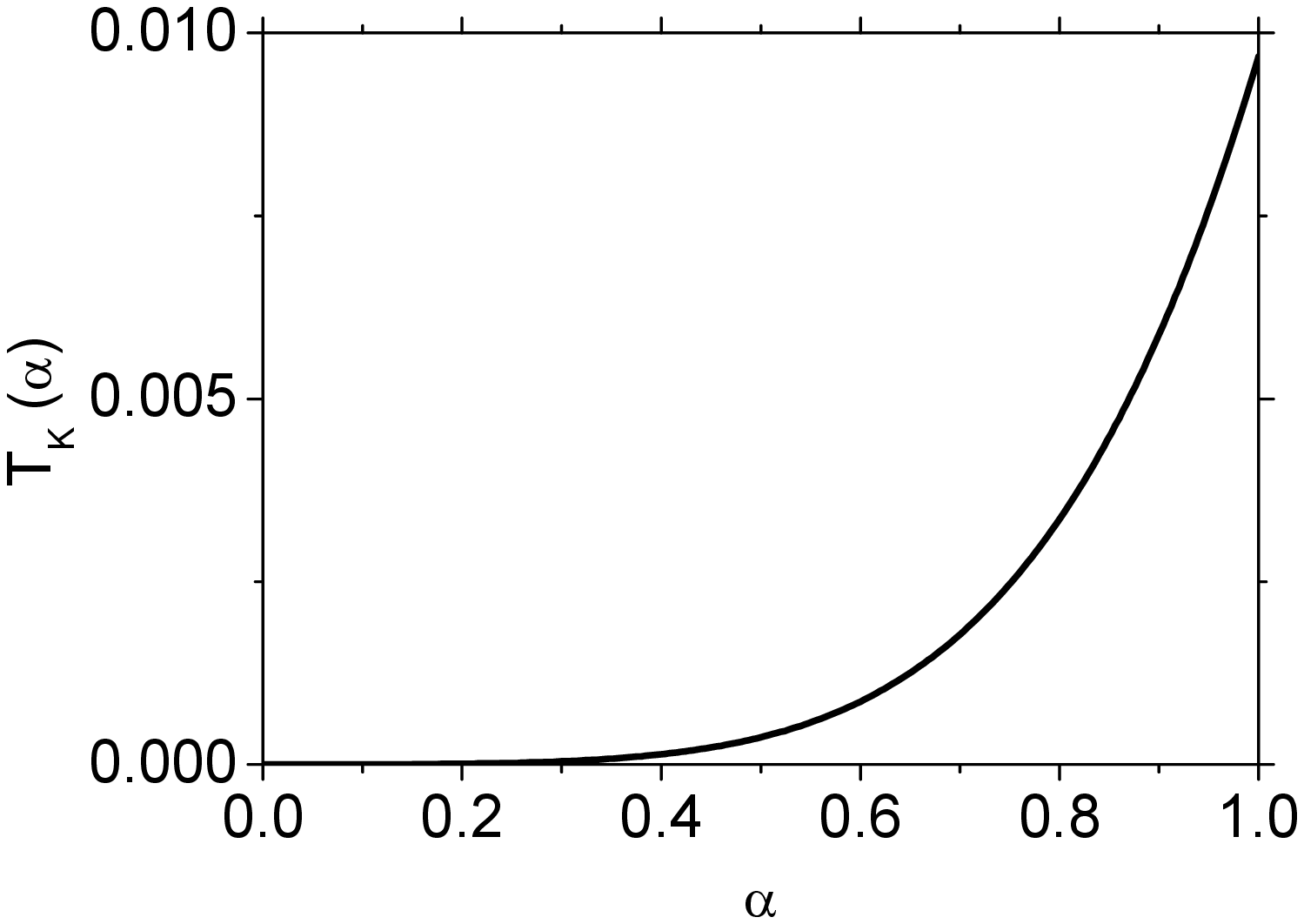}
  \caption{\label{Fig:figTKalfa}
 The Kondo temperature calculated as a function of the asymmetry parameter $\alpha$
 for $q=0$ and for $t=0$.}
  \end{center}
\end{figure}

On the other side, for $q=0$ the following formula describes the
transmission probability,
\begin{equation}\label{Eq:Tq0}
    T(\varepsilon)=\frac{1}{2}\left[\frac{\tilde{\Gamma}^2}{\tilde{\Gamma}^2+(\tilde{\varepsilon}_0-\varepsilon)^2}+
    \frac{\alpha^2}{\alpha^2\tilde{\Gamma}^2+(\tilde{\varepsilon}_0-\varepsilon)^2}\right].
\end{equation}
Figure~\ref{Fig:5} illustrates graphically Eq.~(\ref{Eq:Tq0}) for
indicated values of the asymmetry parameter $\alpha$. This plot
clearly shows, as explained earlier, that the width of the Kondo
peak decreases as $\alpha$ is reduced. This case resembles the
spin Kondo effect in single-level quantum dot coupled to
ferromagnetic leads~\cite{martinekPRL03}. Then the asymmetry
parameter $\alpha$ can be assigned with lead's
(pseudo)polarization in the following way
$\tilde{p}=(1-\alpha)/(1+\alpha)$~\cite{trochaPRB10}. Increasing
pseudopolarization (decreasing $\alpha$) the Kondo effect is
suppressed which is in agreement with
Refs.~\onlinecite{trochaPRB10,martinekPRL03}. In the presence of
asymmetry in couplings ($\alpha<1$) the Kondo temperature should
be defined by geometric mean as:
$T_K\equiv\sqrt{\tilde{\Gamma}_1\tilde{\Gamma}_2}=\tilde{\Gamma}\sqrt{\alpha}$
(remembering that $\tilde{\varepsilon}_0\rightarrow 0$ in deep Kondo regime).
Fig.~\ref{Fig:figTKalfa} shows $\alpha$ dependence of the Kondo
temperature for case of $q=0$.
%
In the present case, $q=0$, electrons may travel only directly
through dot QD1 or QD2 thus the reduction of the Kondo temperature with
decreasing $\alpha$ clearly reflects the suppression of Kondo
fluctuations in the DQD (due to reduction of the charge exchange
between the dots) as one of the dot becomes detached from the
leads and Kondo effect diminishes .

\begin{figure}[t]
\begin{center}
  \includegraphics[width=0.8\columnwidth]{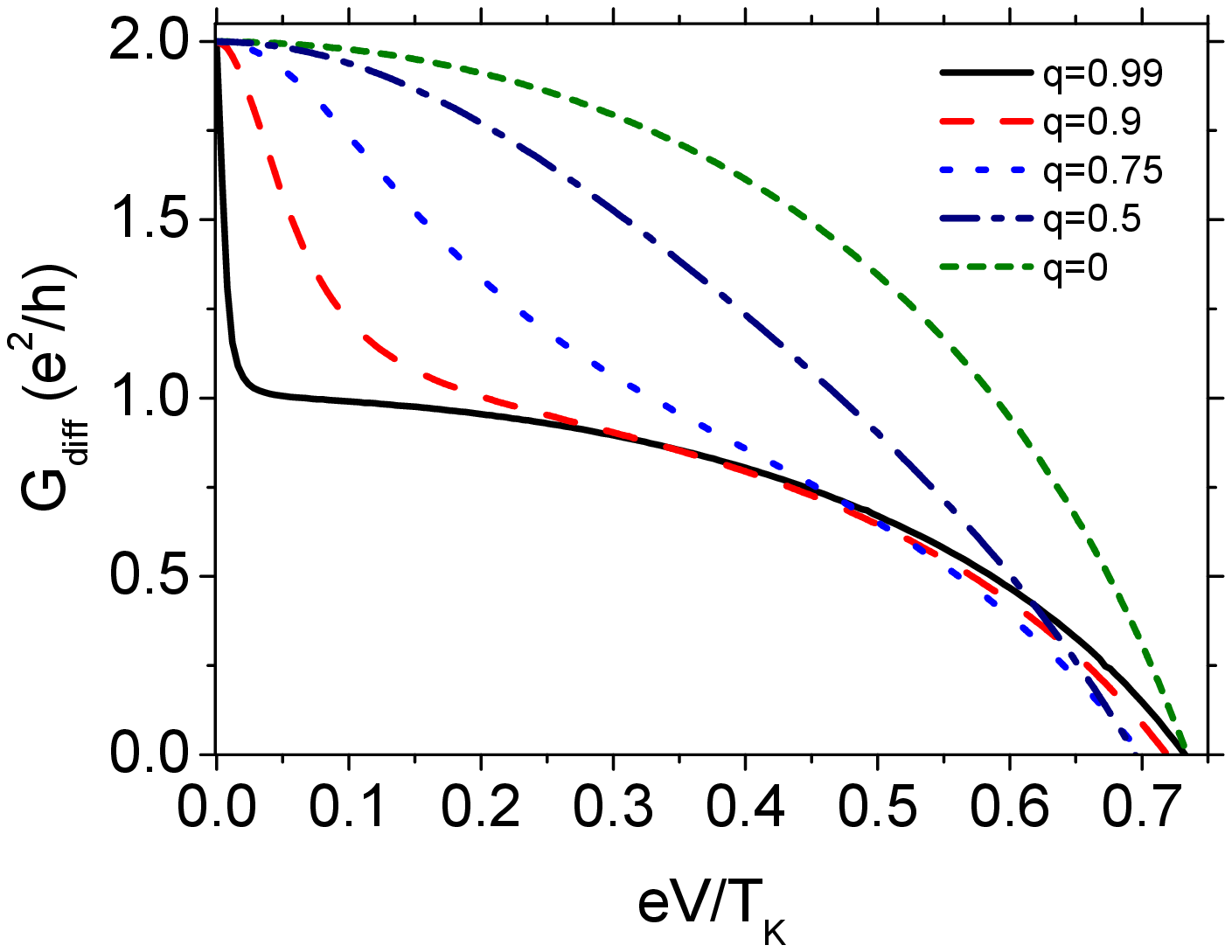}
  \caption{\label{Fig:biasqp}
Differential conductance as a function of bias voltage calculated
for indicated values of parameter $q$ ($q_L=q_R=q$) and for
$\alpha=1$, $t=0$.}
  \end{center}
\end{figure}

Similar dependance on parameter $q$ can be noticed in the
differential conductance which is displayed in
Fig.~\ref{Fig:biasqp}. The differential conductance is symmetric
in respect to the zero bias point, thus we plot only its bias
dependance for nonnegative bias voltage. For $q$ close to unity
the differential conductance acquires Dicke line shape. At zero
bias the differential conductance approaches unitary limit
($2e^2/h$) for all $q<1$. For $q$ close to $1$ the differential
conductance drops very fast to the half of the zero bias value
with increasing the bias voltage and next decreases slowly with
further increasing of the bias voltage. Such a sudden drop of the
differential conductance is absent for smaller values of the
parameter $q$. This feature in the differential conductance is
related to Dicke resonance.

%
\begin{figure}[t]
\begin{center}
  \includegraphics[width=0.92\columnwidth]{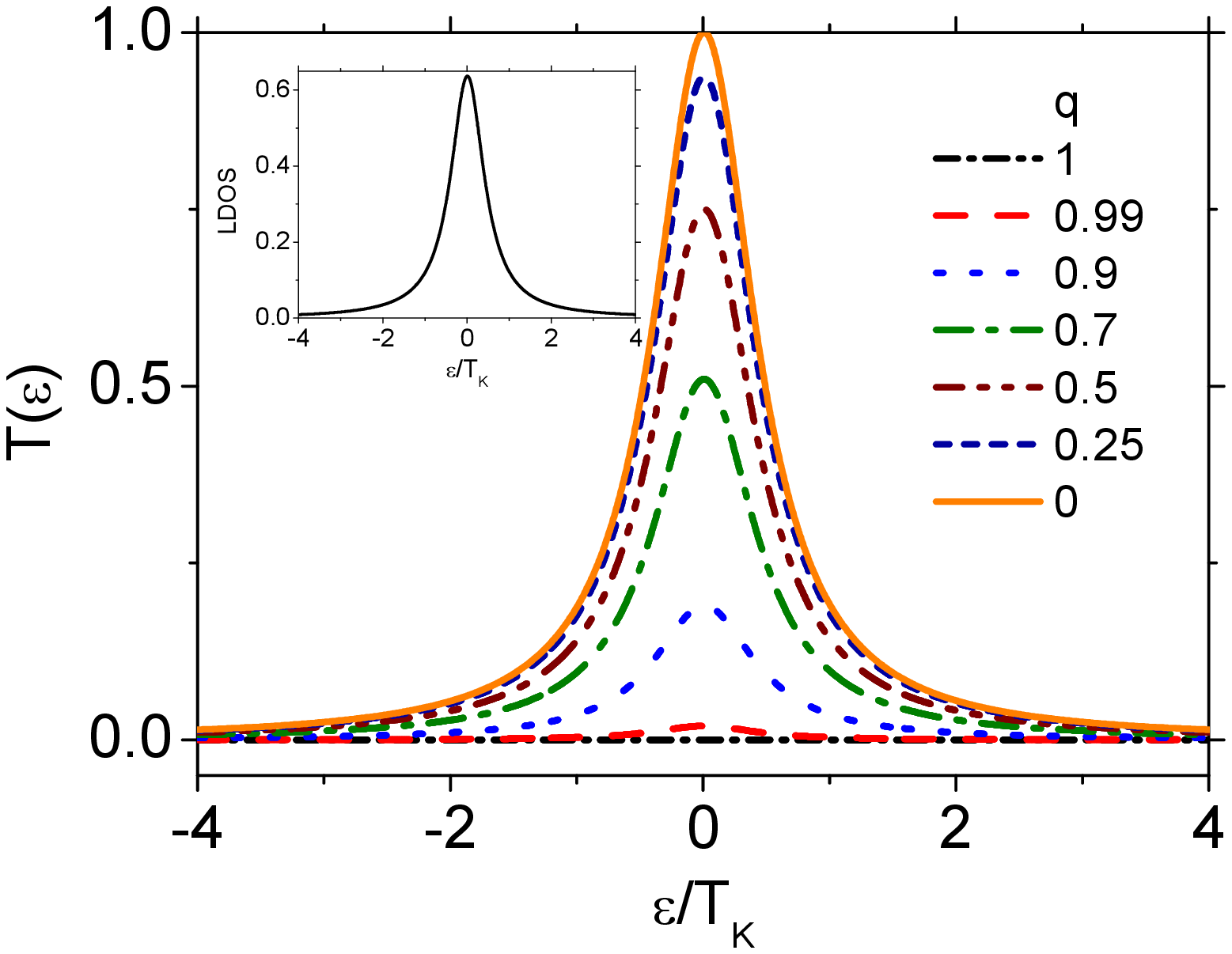}
  \caption{\label{Fig:qm}
  The transmission coefficient calculated for indicated values of the
  parameter $q$ ($q_L=-q_R=q$) and for $\alpha=1$. The inset show corresponding LDOS which
  is independent on the value of $q$.}
  \end{center}
\end{figure}

As we mentioned in Sec.~\ref{Sec:1} the off--diagonal elements of
the coupling matrix can have opposite signs. Specifically, we
examine the case $q_L=q$ and $q_R=-q$. At the beginning we assume
symmetric couplings, i.e., $\alpha=1$. Then for maximal value of
the parameter $q$ ($q=1$) each DQD's molecular state couples to
different reservoirs and as there is no connection between them,
the current doesn't flow. This effect originates from the totally
destructive quantum interference which leads to vanishing of the
transmission even if the LDOS of each dots is finite. It is wort
noting that both LDOS and the corresponding Kondo temperature do
not depend on the parameter $q$, as the contributions from
nondiagonal processes, refereing to the left $L$ and right $R$
lead, cancel each others. This implies that the presented effect
originates fully from the quantum interference. One can notice
that decreasing the value of the parameter $q$, the transmission
can be recovered as is shown in Fig.~\ref{Fig:qm}. Finally, for
$q=0$ the maximal value of the transmission is restored. This is
because as the value of the off-diagonal matrix elements are
decreased the destructive interference becomes totally suppressed,
and thus, for zero value of the nondiagonal couplings the zero
bias conductance is fully restored. This can be shown more
formally by performing the transformation of the dots operators to
the bonding-like ($d_{b}=1/\sqrt{2}(d_1+d_2)$) and
antibonding-like ($d_{a}=1/\sqrt{2}(d_1-d_2)$) state. The
corresponding couplings to the left lead acquires form
$\tilde{\Gamma}_{b,a}^L=\frac{1}{2}(\tilde{\Gamma}_{11}^L+\tilde{\Gamma}_{22}^L)\pm
q\sqrt{\tilde{\Gamma}_{11}^L\tilde{\Gamma}_{22}^L}$, whereas
couplings to the right electrode is given by
$\tilde{\Gamma}_{b,a}^R=\frac{1}{2}(\tilde{\Gamma}_{11}^R+\tilde{\Gamma}_{22}^R)\mp
q\sqrt{\tilde{\Gamma}_{11}^R\tilde{\Gamma}_{22}^R}$. Now, it is
clear that for $\alpha=1$ and $q=1$ the bonding state is coupled
to (decoupled from) left (right) lead, whereas antibonding state
is coupled to (decoupled from) right (left) electrode. As the
parameter $q$ becomes less than $1$, the decoupled states from
given leads for $q=1$ start to bound with them. As a result the
transmission grows with decreasing the value of the parameter $q$.

\begin{figure}[t]
\begin{center}
  \includegraphics[width=0.74\columnwidth]{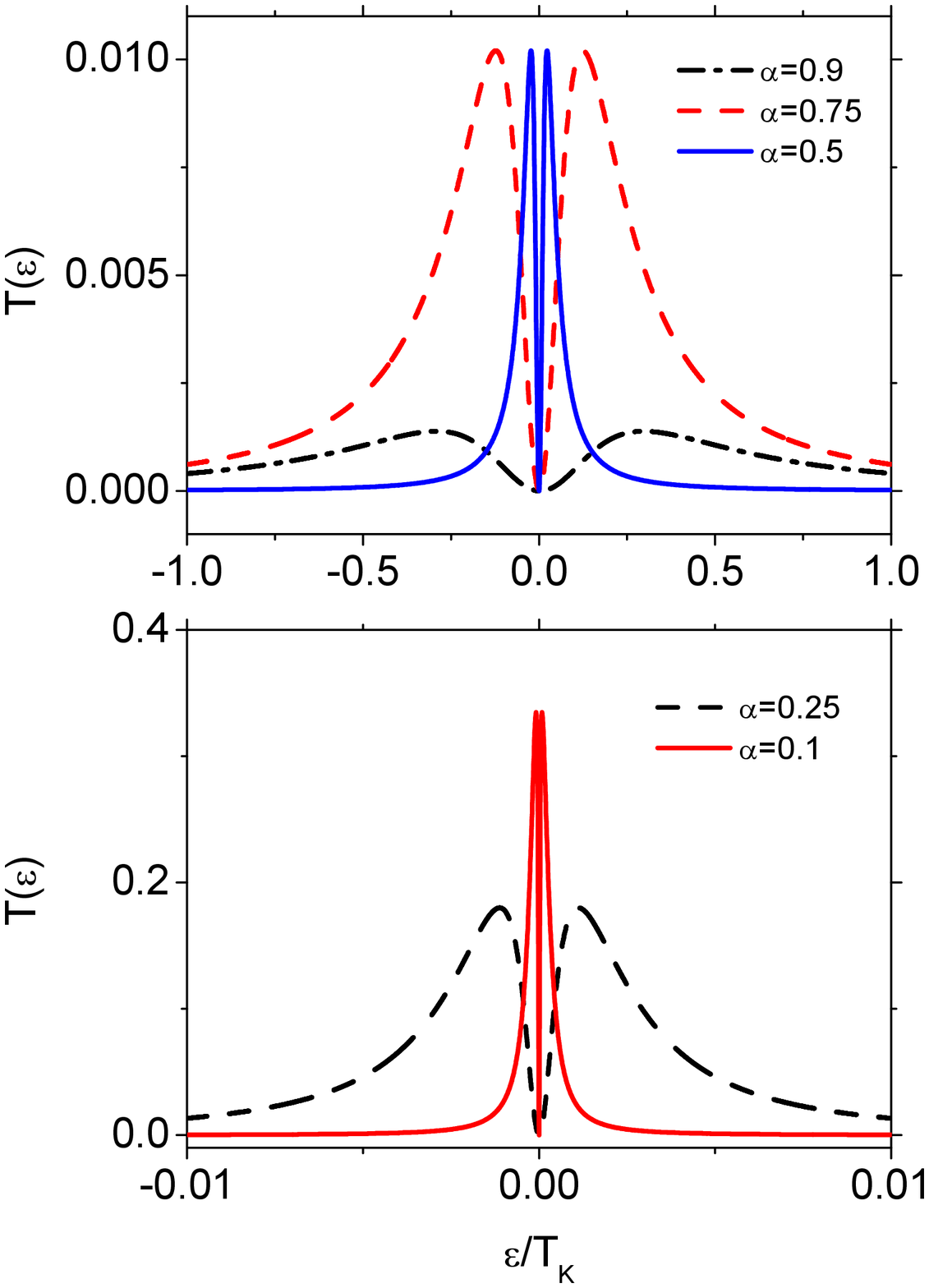}
  \caption{\label{Fig:Tqmalfa}
  The transmission coefficient calculated for indicated values of the parameter
  $\alpha$ and for $q_L=-q_R=1$.}
  \end{center}
\end{figure}

\begin{figure}[t]
\begin{center}
  \includegraphics[width=0.9\columnwidth]{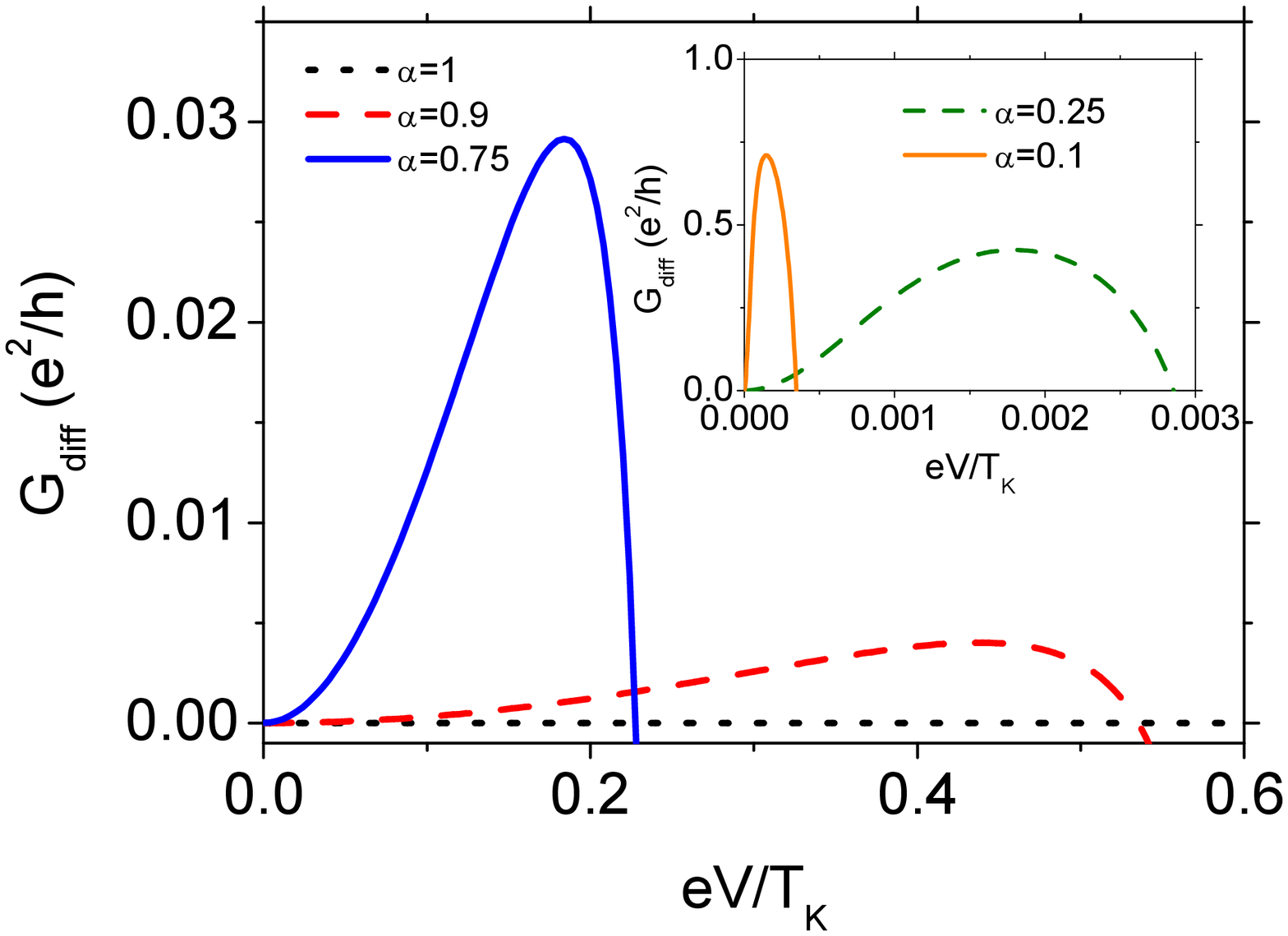}
  \caption{\label{Fig:figbiasqm}
  Differential conductance as a function of bias voltage calculated for indicated values of the parameter
  $\alpha$ and for $q_L=-q_R=1$, $t=0$.}
  \end{center}
\end{figure}

Another interesting quantum interference effect can be found
changing the asymmetry in couplings of two dots to the leads. At
the beginning we keep $q$ equal to $1$ and change the parameter
$\alpha$ in the interval $\langle 0,1\rangle$. As before the
transmission becomes recovered as the asymmetry increases [see
Fig.~\ref{Fig:Tqmalfa}]. However, in comparison to the previous
case, a new feature emerges in the transmission. More
specifically, the dip structure appears in the vicinity of the
zero energy. At $\varepsilon=0$ the transmission drops to zero
which results in vanishing of the linear conductance. Thus, the
zero bias Kondo anomaly is totally
suppressed for any value of the
asymmetry parameter $\alpha$. To verify this effect experimentally
it is desired to measure the differential conductance as a
function of bias voltage. In Fig.~\ref{Fig:figbiasqm} we displayed
bias voltage dependence of the differential conductance for
indicated values of the asymmetry parameter $\alpha$. These
results show that applying finite bias voltage, the differential
conductance (which is zero at zero bias for all $\alpha$) is
restored. The differential conductance becomes the most pronounced
when one of the dot is almost decoupled from the leads, i.e., as
$\alpha\rightarrow 0$, the differential conductance tend to
unitary limit. However, with increasing asymmetry in the couplings
(decreasing $\alpha$) the range of finite values of the
differential conductance shrinks due to decreasing of the Kondo temperature. It is worth noting that
the corresponding Friedel phase is a continuous function of energy -- it
does not suffer discontinuity at $\varepsilon=0$ because no BIC
appears. However, the transmission at $\varepsilon=0$ vanishes,
thus the phase of transmission amplitude should be discontinuous
at $\varepsilon=0$.
%

\subsection{Fano effect}\label{Sec:3B}
\begin{figure}[t]
\begin{center}
  \includegraphics[width=0.82\columnwidth]{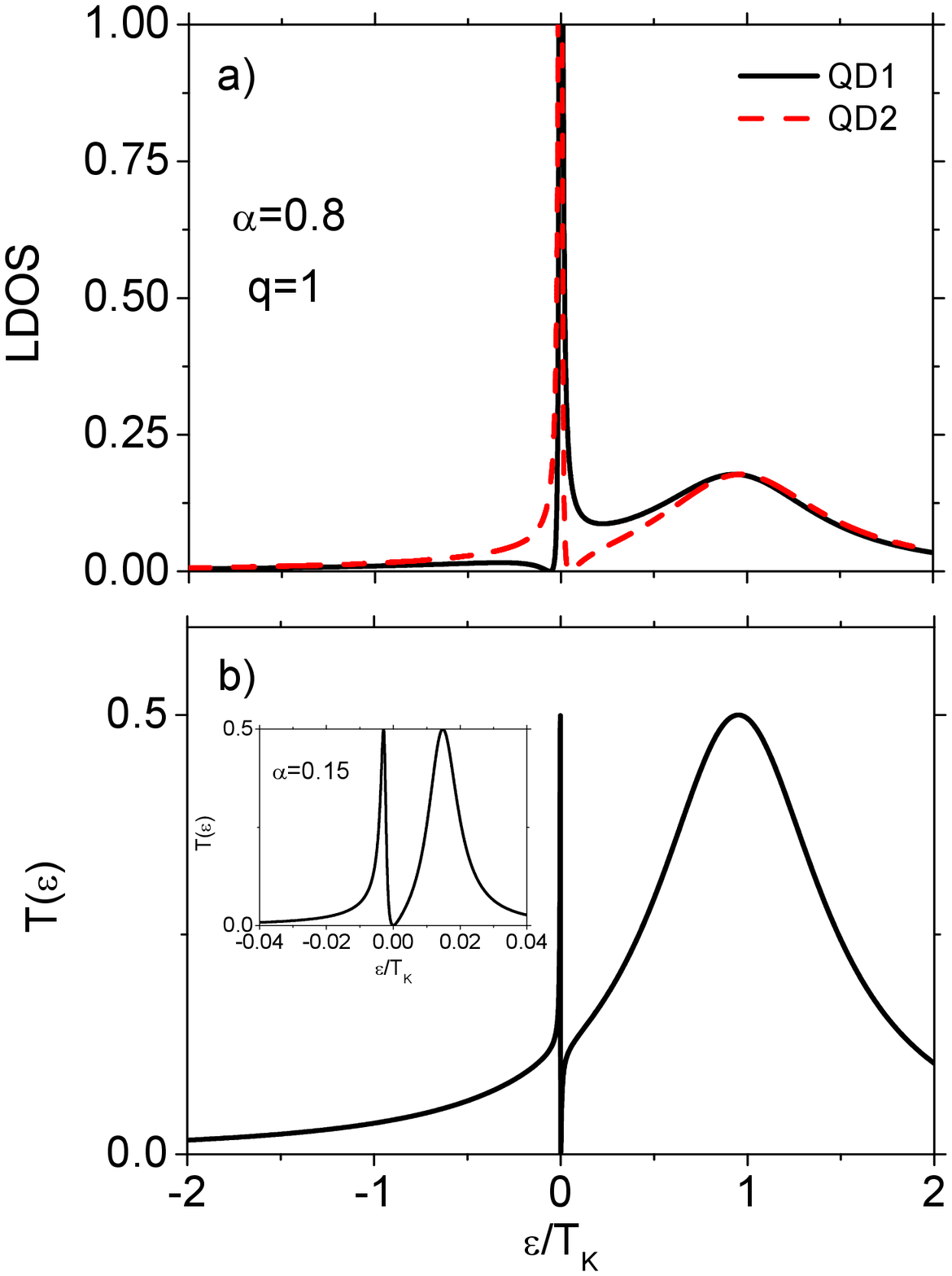}
  \caption{\label{Fig:6}
Local density of states for the both dots (a) and the transmission
  probability (b) calculated for $q=1$,
  $\alpha=0.8$ and for $t=0.8$. The existence of the Fano-Kondo effect is well
  visible.
  The inset shows the transmission probability obtained for $q=1$ and for $\alpha=0.15$.}
 \end{center}
  \end{figure}

Here, we show the results obtained for nonzero interdot hopping
parameter ($t\neq 0$). In present situation the interdot hopping
term lifts the degeneracy and the bonding and the antibonding
levels are split away. As a result, the density of states of the
DQD system coupled to the leads consists of two Kondo peaks; a
broad peak centered at the bonding state and a narrow one
corresponding to the antibonding state. This is clearly showed in
Fig.~\ref{Fig:6}(a)) where the LDOS for both dots is plotted for
maximal off-diagonal matrix elements ($q=1$). It is very
interesting that the narrow peaks in LDOS for QD1 and QD2 have
opposite symmetries. The LDOS for QD1 reaches zero for negative
energy, whereas the one for QD2 comes to zero for positive energy.
Usually, when the asymmetry in coupling of two dots to the leads
is reduced the width of bonding (antibonding) state increases
(decreases). Finally, for $\alpha=1$ (full symmetric system) the
antibonding states is decoupled from the leads and acquires
$\delta$-Dirac shape (it becomes a BIC), whereas the bonding state
acquires the width of $2\tilde{\Gamma}$. On the other side, when
the asymmetry in coupling of two dots to the leads is increased
then the width of bonding (antibonding) state decreases
(increases) and for $\alpha=0$ the two resonances acquire the same
width. However, in current problem the situation is more complex
because the level's widths are renormalized by factor
$\tilde{b}^2$ which has to be obtained self-consistently for each
$\alpha$. Now it is only true that relative width of the bonding
state to the width of the antibonding state grows as $\alpha$
increases, and are the same for $\alpha=0$. However, due to widths
renormalization both peaks in LDOS may have smaller widths for
small $\alpha$ with comparison to the width of the narrow peak for
large $\alpha$ (but not $\alpha\approx 1$) (for instance for
$\alpha=0.15$ both resonances are merged into narrow resonance for
$\alpha=0.8$). Similar behavior is observed in the LDOS for dots
QD1 and QD2.
 Another feature is that for smaller values of the $\alpha$
only LDOS for QD2 may reach zero value at some energy.
  %
\begin{figure}[t]
\begin{center}
  \includegraphics[width=0.65\columnwidth]{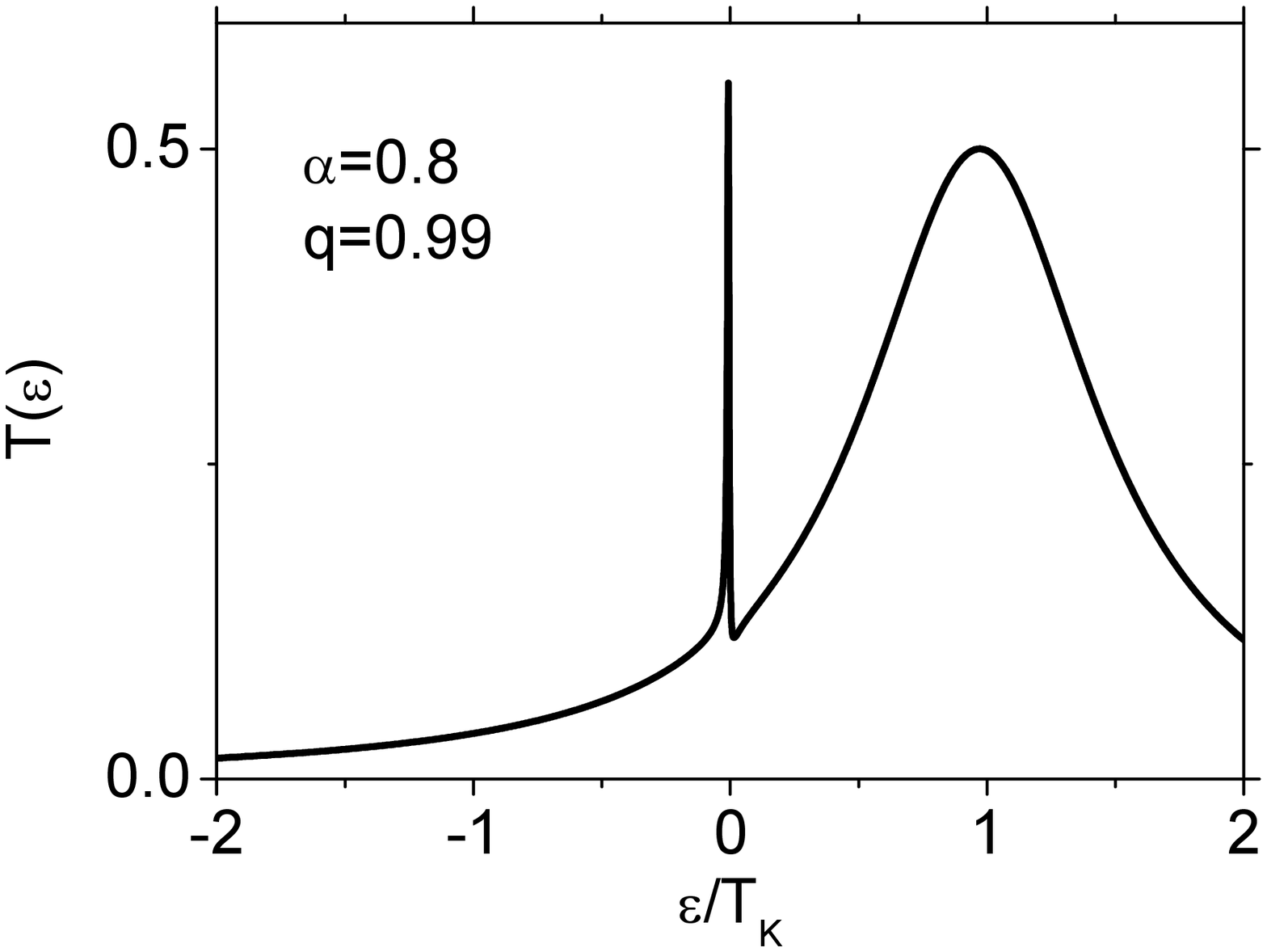}
  \caption{\label{Fig:7}
 The transmission probability calculated for $\alpha=0.8$, $t=0.8$, and for
  $q=0.99$. Even a small shift in $q$ leads to the considerably
  changes in the transmission probability within the antibonding level
considerably (compare with Fig.~\ref{Fig:6}).}
  \end{center}
  \end{figure}
The above behavior of LDOS for $\alpha\in (0,1)$ results in Fano
antiresonance in the transmission probability. The corresponding
transmission probability calculated for $q=1$ and $\alpha=0.8$ is
displayed in Fig.~\ref{Fig:6}(b). One can notice that 
the transmission reveals
the antiresonance character with a characteristic Fano line shape.
In turn, the transmission function corresponding to the bonding state is
relatively broad and (roughly) Lorenzian. In inset of
Fig.~\ref{Fig:6}(b) the $T(\varepsilon)$ is plotted for greater
asymmetry in coupling of two dots to the leads ($\alpha=0.15$). It
is noticed that the antiresonance behavior is still preserved but
now the widths' ratio of the narrow resonance with respect to the
broad one is much larger in comparison to the previous case (i.e.,
when $\alpha=0.8$). With decreasing $\alpha$ the widths of both
peaks also decrease which can be seen looking at the energy scales
for both cases. This means that the Kondo temperature is also
lowered as $\alpha$ decreases (as was shown in the previous section). For $q=1$ the transmission
probability has poles at,
\begin{equation}\label{Eq:poles}
    \varepsilon=\tilde{\varepsilon}_0-\frac{2\sqrt{\alpha}}{1+\alpha}\tilde{t}.
\end{equation}
It is worth to mention that at this energy the phase of the transmission amplitude suffers discontinuity.
This is no more true for $q<1$, because then the transmission
probability has complex poles. In turn, for $q<1$ the transmission
probability does not reach zero and for specific value of $q$
antiresonance behavior is less visible. Even a small shift in $q$
change the transmission probability within the antibonding level
considerably, which is showed in Fig.~\ref{Fig:7}.

At this point we should remark that the slave boson mean
field approach does not take into account the level
renormalization arising due to coupling of dots to the leads. Such
a renormalization should lead to splitting of the zero bias
anomaly for specific cases.~\cite{trochaPRB10} The level splitting
can occurs due to asymmetry in coupling of the dots to the leads,
i.e., for $\alpha<1$ as has been shown in
Ref.~\onlinecite{trochaPRB10} by means of the scaling technique.
However, such a splitting can also be induced by indirect coupling
of the dots. One can show that the splitting is proportional to
the strength of the indirect coupling. Moreover, the level
renormalization of a given state is proportional to the coupling
strength of the other level. This implies that for $q$ close to $1$
the bonding-like level will be only  weakly renormalized, whereas
the antibonding-like level will experience strong renormalization.

Thus, the transmission described in Sec.~\ref{Sec:3A} should
resembles that from Fig.~\ref{Fig:7} but with the broad peak pinned
close to $\varepsilon=0$ and the narrow maximum shifted up in
energy for relatively large value of parameter $q$.
One can show this more formally including corrections in
molecular-like levels due to mentioned renormalization.

\begin{figure}[t]
\begin{center}
  \includegraphics[width=0.6\columnwidth]{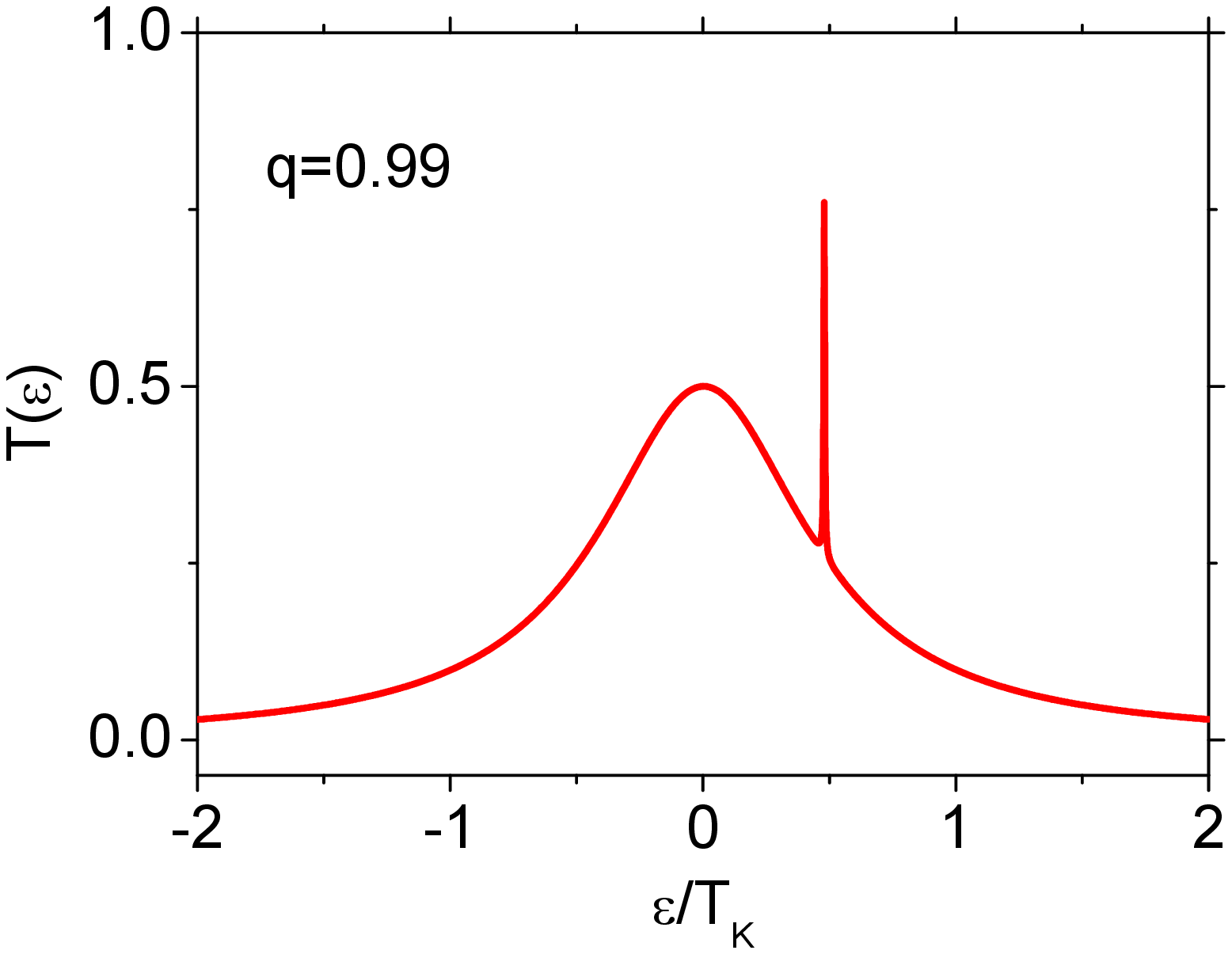}
  \caption{\label{Fig:figx}
  The transmission coefficient including the level renormalization calculated for
  $\alpha=1$ and for $q_L=q_R=0.99$.}
  \end{center}
\end{figure}

To correct the drawback of the SBMF method one can by hand
introduce the mentioned renormalization of the levels as follows:
$\varepsilon_b=\varepsilon_0+\delta\varepsilon_b$ and
$\varepsilon_a=\varepsilon_0+\delta\varepsilon_a$ (with $t=0$ for the sake of simplicity). Here,
$\delta\varepsilon_i$ are the corrections due to indirect coupling
of the dots levels. These corrections should be determined using
relevant technique, as for instance earlier mentioned scaling
procedure. However, as the SBMF technique fails for nondegenerate
states, one can not introduce by hand the renormalized levels into
self-consistent equations of the form (\ref{Eq:SC1}) and
(\ref{Eq:SC2}) derived within bonding and antibonding states
basis. Thus, we give here only rough estimation and some
predictions implying from the levels scaling. Such approach should
deliver qualitatively good insight into Kondo peak splitting
phenomenon, however to obtain quantitatively consistent results
more reliable technique should be applied for the considered
problem. For the sake of simplicity we analyze only the symmetric
case ($\alpha=1$). Introducing the level renormalization the
transmission acquires the following form:
\begin{equation}\label{Eq:TransmEO}
    T(\varepsilon)=\frac{\tilde{\Gamma}^2(1-q)^2}{(\varepsilon-\tilde{\varepsilon}_a)^2+
    \tilde{\Gamma}^2(1-q)^2}
    +
    \frac{\tilde{\Gamma}^2(1+q)^2}{(\varepsilon-\tilde{\varepsilon}_b)^2+
    \tilde{\Gamma}^2(1+q)^2}
\end{equation}
with renormalized bonding ($b$) and antibonding ($a$) levels of the following assumed
form~\cite{trochaUnpub}:
\begin{equation}\label{Eq:levelsR}
 \varepsilon_i=\varepsilon_0+\Gamma_{\bar{i}}\Delta
\end{equation}
 Here, $\Delta$~\cite{trochaApp} stands for some function which in general depends on the system's parameters (like bandwidth, dots energy levels, couplings). For the sake of simplicity we assume $\Delta$ to be a constant number and being maximum value of function $\Delta (q)$ obtained from the scaling procedure, i.e. when $q=1$. Equation (\ref{Eq:TransmEO}) clearly shows that Kondo peak becomes split due to the level renormalization originating from indirect tunneling between the dots. In Fig.~\ref{Fig:figx} we display expected lineshape in the transmission for relatively large value of parameter $q$. The splitting in the transmission should decrease with decreasing the value of parameter $q$ and beyond a certain value of $q$ the splitting ceases to be visible. For $q=0$ (and $\alpha=1$) no splitting occurs.

%


\section{Summary and conclusions}

In this paper we have investigated electronic properties of double
quantum dots coupled to external leads. The dots have been coupled
both via hopping term and Coulomb interaction. Moreover, we have
considered also the effects of indirect tunneling between the dots
through the leads. Employing the slave-boson mean field approach
the local density of states for both dots, the Friedel phase and 
the transmission in Kondo regime have been calculated. Moreover, 
to be more familiar with experiment we have calculated the 
corresponding differential conductance.

We have shown that for some set of parameters the Dicke-and
Fano-like resonances may appear in the considered system. More
specifically, it has been noticed that for zero interdot hopping
parameter $t$ the LDOS of each QD consist of broad and narrow
Kondo peaks that are superimposed. As in original Dicke effect one
may associate narrow (broad) central peak in LDOS with a
subradiant (superradiant) mode. Moreover, Dicke line shape has
been found in the transmission and in the differential conductance.
We have observed that this
effect is very sensitive to the strength of the off-diagonal
matrix elements; with reducing the value of the off-diagonal
matrix elements the Dicke line  both in LDOS and the transmission
is transformed into usual Lorenzian line. It has been found
that when the interdot tunneling is allowed the transmission
probability may reveal the antiresonance behavior with a
characteristic Fano line shape. Moreover, we have noticed that the
line shape of the antiresonance is also very susceptible to the change
of the value of the off-diagonal matrix elements.

We have also calculated the Kondo temperature, and also have shown that
the latter becomes suppressed  with increasing  asymmetry in the
dot-lead couplings when there is no indirect coupling. Moreover,
when the indirect coupling is turned on, the characteristic widths
for both distinct channels behave in different way with varying
strength of the off-diagonal tunneling processes. We found also that the corresponding Kondo temperature is
totally suppressed for maximal value of the
indirect coupling and no Kondo effect occurs.
Moreover, we have also included level renormalization effects due to indirect coupling phenomenon, which leads to the splitting of the Kondo peak.


\begin{acknowledgements}
 The author thanks prof. J\'ozef Barna\'s for constructive criticism
 and fruitful discussions.
This work was supported by Ministry of Science and Higher
Education as a research project N N202 169536 in years 2009-2011.
\end{acknowledgements}

\appendix
\section{Proof of $\tilde{\varepsilon}_0\rightarrow 0$ in deep Kondo regime}
Here, we show that in deep Kondo regime the renormalized
parameter $\tilde{\varepsilon}_0$ is equal to zero and, thus, the
Kondo temperature strictly correspond to to the renormalized width
$\tilde{\Gamma}_{}$ of the Abricolov-Suhl resonance. This can be
shown analytically by integrating self-consistent equations for
slave-boson parameters, $\tilde{b}$, $\lambda$,  written in
representation of {\it bonding} and {\it antibonding} states.

In the basis of the {\it bonding} and {\it antibonding} states the
Hamiltonian of the system becomes diagonal for
$\varepsilon_1=\varepsilon_2\equiv\varepsilon_0$ and acquires the
following form:
\begin{eqnarray}\label{Eq:HamiltonianEO}
   \hat{H}=\hat{H}_c&+&\sum_{i=b,a}\limits\varepsilon_{i}d^\dag_{i}d_{i}+
   Un_{b}n_{a}
   \nonumber \\
    &+&\sum_{\mathbf{k}\beta}\limits\sum_{i=e,o}
   \limits (V_{i\mathbf{k}}^\beta c^\dag_{\mathbf{k}\beta}d_{i}+\rm
   H.c.),
\end{eqnarray}
with $\varepsilon_b=\varepsilon_0+t$,
$\varepsilon_a=\varepsilon_0-t$. In further considerations we
assume $t=0$, thus, $\varepsilon_b=\varepsilon_a=\varepsilon_0$.

In the mean field slave boson representation the Hamiltonian
(\ref{Eq:HamiltonianEO}) acquires the following form:
\begin{eqnarray}\label{Eq:HamiltonianEOSB}
\tilde{H}^{MF}=\hat{H}_c&+&
\sum_{i=\emph{b,a}}\limits\tilde{\varepsilon}_{0}f^\dag_{i}f_{i} +
\lambda\left(\tilde{b}^2-1\right)
  \nonumber \\
   &+&   \sum_{{\mathbf k}\beta}\sum_{i=\emph{b,a}}\limits
   (\tilde{V}_{i\mathbf{k}}^\beta
   c^\dag_{{\mathbf k}\beta}f_{i}+{\rm H.c.}).
\end{eqnarray}

The self-consistent equations determining the unknown parameters
$\tilde{b}$ and $\lambda$ have the form of Eqs.~(\ref{Eq:SC1}) and
(\ref{Eq:SC2}) with lesser Green's function $\langle\langle
f_{i}|f^{\dag}_{j}\rangle\rangle^<_\varepsilon$ defined in the
basis of the {\it bonding} and {\it antibonding} states. The
lesser {\it bonding} and {\it antibonding}  Green's function has
the following form:
\begin{eqnarray}\label{Eq:Gless}
G_{ii}^{<}(\varepsilon)=\frac{f(\varepsilon)\tilde{\Gamma}_i}
{(\varepsilon-\tilde{\varepsilon}_0)^2+(\tilde{\Gamma}_i/2)^2}.
\end{eqnarray}
At $T=0$ K $f(x)=\theta (-x)$ and the integration of the
self-consistent equations is straightforward~\cite{integrand}
leading to the following expressions:
\begin{eqnarray}\label{Eq:SCba}
\frac{1}{\pi}\sum_{i=b,a}\limits{\rm Im}
\left[\ln\left(\frac{\varepsilon_0+i\tilde{\Gamma}_i/2}{D}\right)\right]=
1-\tilde{b}^2,
\nonumber\\
\frac{1}{2\pi}\sum_{i=b,a}\limits\tilde{\Gamma}_i{\rm Re}
\left[\ln\left(\frac{\varepsilon_0+i\tilde{\Gamma}_i/2}{D}\right)\right]+
\lambda\tilde{b}^2=0.
\end{eqnarray}
In the deep Kondo regime we can approximate: $1-\tilde{b}^2\approx
1$ and $\lambda\approx-\varepsilon_0$ and then
Eqs.~(\ref{Eq:SCba}) simplify to the following equations:
\begin{eqnarray}\label{Eq:SCbaApprox}
&&\sum_{i=b,a}\limits{\rm Im}
\left[\ln\left(\frac{\varepsilon_0+i\tilde{\Gamma}_i/2}{D}\right)\right]=
\pi,
\nonumber\\
&&\sum_{i=b,a}\limits\Gamma_i{\rm Re}
\left[\ln\left(\frac{\varepsilon_0+i\tilde{\Gamma}_i/2}{D}\right)\right]=2\pi\varepsilon_0.
\end{eqnarray}
Here, we  consider only the symmetric case, $\alpha=1$, however,
extension to arbitrary $\alpha$ is straightforward.
For $\alpha=1$ and $q_L=q_R=q$, the couplings to the {\it bonding}
and {\it antibonding} states acquires the following form,
$\Gamma_{b,a}=(1\pm q)\Gamma$. Combining two
equations~(\ref{Eq:SCbaApprox}) one arrives with equation:
\begin{widetext}
\begin{equation}\label{Eq:combine}
(\tilde{\varepsilon}_0+i\tilde{\Gamma}_b/2)(\tilde{\varepsilon}_0+i\tilde{\Gamma}_a/2)
\left(\frac{\tilde{\varepsilon}_0+(\tilde{\Gamma}_b/2)^2}{\tilde{\varepsilon}_0+(\tilde{\Gamma}_a/2)^2}\right)^{\frac{q}{2}}
=D^2\exp\left[\left(\frac{i}{2}+\frac{\varepsilon_0}{\Gamma}\right)2\pi\right],
\end{equation}
\end{widetext}
which real and imaginary parts satisfy the following equalities:
\begin{widetext}
\begin{eqnarray}\label{Eq:combine}
\left(\tilde{\varepsilon}_0^2+\frac{1}{4}\tilde{\Gamma}_b\tilde{\Gamma}_a\right)
\left(\frac{\tilde{\varepsilon}_0+(\frac{\tilde{\Gamma}_b}{2})^2}
{\tilde{\varepsilon}_0+(\frac{\tilde{\Gamma}_a}{2})^2}\right)
^{\frac{q}{2}}&=&-D^2\exp\left(\frac{2\pi\varepsilon_0}{\Gamma}\right),\label{Eq:Real}
\\
\tilde{\varepsilon}_0(\tilde{\Gamma}_b+\tilde{\Gamma}_a)
\left(\frac{\tilde{\varepsilon}_0+(\frac{\tilde{\Gamma}_b}{2})^2}
{\tilde{\varepsilon}_0+(\frac{\tilde{\Gamma}_a}{2})^2}\right)^{\frac{q}{2}}
&=&0\label{Eq:Imaginary}.
\end{eqnarray}
\end{widetext}
Equation~(\ref{Eq:Imaginary}) is satisfied if
$\tilde{\varepsilon}_0=0$ or $\tilde{b}^2=0$. However, the later
solution leads to nonphysical value for $\tilde{\varepsilon}_0$,
thus, the only solution must be $\tilde{\varepsilon}_0=0$. This
result clearly shows that in deep Kondo regime slave-boson
parameter $\tilde{\varepsilon}_0$ vanishes, {\it Q.E.D.}

Introducing the solution for $\tilde{\varepsilon}_0$ into
Eq.~(\ref{Eq:Real}) one finds,
\begin{equation}\label{Eq:b}
\tilde{b}^2=\frac{2D}{\Gamma}\frac{(1-q)^{\frac{q-1}{2}}}{(1+q)^{\frac{q+1}{2}}}
\exp\left(\frac{\pi\varepsilon_0}{\Gamma}\right).
\end{equation}
It is worth noting that the above equation does not determine the
value of $\tilde{b}^2$ for $q=1$ when the SBMF method fails.


\begin{thebibliography}{65}
\bibitem{integrand} At finite temperature the integrands appearing
in the self-consistent equations can be evaluated analitically by
means of contour integration. To satisfy Jordan's lemma the
integrand appearing in the equation of form~(\ref{Eq:SC2}) must be
calculated with the help of lorenzian cut-off function,
$D^2/(\varepsilon^2+D^2)$. Then, the results can be expressed by
means of digamma function and Fermi-Dirac function, which for
$T=0$ converge to the expressions derived in the Appendix.
\end{thebibliography}

\end{document}